\documentclass[11pt,a4paper]{article}
\pdfoutput=1
\usepackage{jheppub}
\input{psfig}
\usepackage{dcolumn}   
\usepackage{subfig}
\usepackage{float}
\newcommand\ltap{\
  \raise.3ex\hbox{$<$\kern-.75em\lower1ex\hbox{$\sim$}}\ }
\newcommand\gtap{\
  \raise.3ex\hbox{$>$\kern-.75em\lower1ex\hbox{$\sim$}}\ }

\newcommand\simge{\mathrel{%
   \rlap{\raise 0.511ex \hbox{$>$}}{\lower 0.511ex \hbox{$\sim$}}}}
\newcommand\simle{\mathrel{
   \rlap{\raise 0.511ex \hbox{$<$}}{\lower 0.511ex \hbox{$\sim$}}}}

\newcommand{\slashchar}[1]%
        {\kern .25em\raise.18ex\hbox{$/$}\kern-.60em #1}
\def\lsim{\mathrel{\raise.3ex\hbox{$<$\kern-.75em\lower1ex\hbox{$\sim$}}}}
\def\gsim{\mathrel{\raise.3ex\hbox{$>$\kern-.75em\lower1ex\hbox{$\sim$}}}}
\newcommand{\bs}{\boldsymbol}

\newcommand\CK{{\cal K}}
\newcommand\CL{{\cal L}}

\newcommand\CO{{\cal O}}

\newcommand\CT{{\cal T}}

\newcommand\be{\begin{equation}}
\newcommand\ee{\end{equation}}
\newcommand\bea{\begin{eqnarray}}
\newcommand\eea{\end{eqnarray}}
\newcommand\ba{\begin{array}}
\newcommand\ea{\end{array}}
\newcommand\nn{\nonumber}
\newcommand\tx{\textstyle}

\newcommand{\thalf}{\textstyle{\frac{1}{2}}}

\newcommand{\tthird}{\textstyle{\frac{1}{3}}}
\newcommand{\tfourth}{\textstyle{\frac{1}{4}}}

\newcommand\gev{{\rm GeV}}
\newcommand\tev{{\rm TeV}}

\newcommand\fb{{\rm fb}}
\newcommand\ifb{{\rm fb}^{-1}}

\newcommand\lvac{\langle \Omega \vert}

\newcommand\ellm{\ell^-}

\newcommand\ellp{\ell^+}

\newcommand\suc{SU(3)_C}
\newcommand\Ntc{N_{TC}}
\newcommand\sutc{SU(N_{TC})}

\newcommand\atc{\alpha_{TC}}

\newcommand\Letc{\Lambda_{ETC}}
\newcommand\Ltc{\Lambda_{TC}}

\title{The light composite Higgs boson in strong Extended Technicolor}
\author[a]{Kenneth Lane,}
\author[a]{Lukas Pritchett}
\affiliation[a]{Department of Physics, Boston University, 590 Commonwealth Avenue, Boston, Massachusetts 02215, USA}
\emailAdd{lane@physics.bu.edu}
\emailAdd{lpritch@bu.edu}

\abstract{ 
  This paper extends an earlier one describing the Higgs boson $H$ as a light
  composite scalar in a strong extended technicolor model of electroweak
  symmetry breaking. The Higgs mass $M_H$ is made much smaller than
  $\Lambda_{ETC}$ by tuning the ETC coupling very close to the critical value
  for electroweak symmetry breaking. The technicolor interaction, neglected
  in the earlier paper, is considered here. Its weakness relative to extended
  technicolor is essential to understanding the lightness of $H$ compared to
  the low-lying spin-one technihadrons. Technicolor cannot be completely
  ignored, but implementing technigluon exchange together with strong
  extended technicolor appears difficult. We propose a solution that turns
  out to leave the results of the earlier paper essentially unchanged. An
  argument is then presented that masses of the spin-one technifermion bound
  states, $\rho_H$ and $a_H$, are much larger than $M_H$ and, plausibly,
  controlled by technicolor. Assuming $M_{\rho_H}$ and $M_{a_H}$ are in the
  TeV-energy region, we identify $\rho_H$ and $a_H$ with the diboson excesses
  observed near $2\,\tev$ by ATLAS and CMS in LHC Run~1 data, and we discuss
  their phenomenology for Runs~2 and~3.
}

		
\begin{document}
\maketitle
\flushbottom

\newpage

\section{Overview}

The discovery at CERN in 2012 of a Higgs boson, $H$, at
$125\,\gev$~\cite{Aad:2012tfa, Chatrchyan:2012ufa} --- consistent so far with
the lone Higgs boson of the standard model --- has made untenable the
original idea of technicolor as the source of electroweak symmetry breaking
(EWSB)~\cite{Weinberg:1979bn, Susskind:1978ms}. In technicolor (TC), the new
boson $H$ is interpreted as the lightest $0^+$ bound state of
technifermions. But, then, one expects other bound technihadrons as light as
this scalar, especially the spin-one vector and axial vector states, and
resonances this light have not been seen.\footnote{Refs.~\cite{Foadi:2012bb,
    Belyaev:2013ida} have argued that the mass of the lightest scalar
  technihadron is greatly reduced from its expected value of 100s of GeV by
  the negative top-loop contribution.  This requires a large top mass from
  extended technicolor and a large coupling of $\bar tt$ to the scalar,
  neither of which are explained in these references. These papers also
  ignore the strong-TC coupling of the scalar to the Goldstone bosons of
  EWSB. A simple model calculation shows that this positive loop-contribution
  overwhelms the negative one from top.}$^,$\footnote{Many papers have
  suggested that the light Higgs boson is a techni-dilaton, i.e., a
  pseudo-Goldstone boson (PGB) of spontaneously broken conformal symmetry in
  walking technicolor. See Refs.~\cite{Yamawaki:1986zg, Goldberger:2008zz,
    Appelquist:2010gy,Bellazzini:2012vz} for a sampling.  This is an
  appealing idea, but it is difficult to understand how such a light PGB can
  arise when the explicit breaking giving rise to its mass, namely
  spontaneous chiral symmetry breaking in technicolor, is as strong as the
  near-conformal dynamics of technicolor itself.}

An attractive dynamical alternative to TC is that $H$ is a composite state
bound by strong interactions active well above the weak scale $\Lambda_{EW}$
of several $100\,\gev$. But, unlike TC, all composite models and, indeed, all
models so far of $H$ require some degree of
fine-tuning~\cite{Contino:2010rs,Giudice:2013yca,Bellazzini:2014yua,
  Barnard:2014tla} to be consistent with ATLAS and CMS
measurements~\cite{Agashe:2014kda} and
searches.\footnote{\tt{https://twiki.cern.ch/twiki/bin/view/AtlasPublic/ExoticsPublicResults}\\
 \indent
 \tt{https://twiki.cern.ch/twiki/bin/view/AtlasPublic/SupersymmetryPublicResults}\\ 
\indent  \tt{https://twiki.cern.ch/twiki/bin/view/CMSPublic/PhysicsResultsEXO}\\
\indent  \tt{https://twiki.cern.ch/twiki/bin/view/CMSPublic/PhysicsResultsSUS}\\
\indent  \tt{https://twiki.cern.ch/twiki/bin/view/CMSPublic/PhysicsResultsB2G}}
This tuning may be loosely characterized by $\Lambda_{EW}^2/\Lambda^2$, where
$\Lambda$ is the physical scale of the new dynamics or particles stabilizing
$M_H$.

Motivated by these considerations, we proposed a new composite model of $H$
employing {\em strong} extended technicolor (ETC) as the main driver of
EWSB~\cite{Lane:2014vca}. As in the standard model (SM) and TC, the fermions
in this model transform as left-handed doublets and right-handed singlets
under $(SU(2)\otimes U(1))_{EW}$. If the ETC interaction's strength exceeds a
critical value, it generates nonzero top quark and technifermion masses,
$m_t$ and $m_T$, thus breaking electroweak (EW) symmetry. Generically, these
masses are of order the ETC scale, $\Lambda = \Letc$, of several
$100\,\tev$. But this can be avoided if the ETC coupling is tuned to within
$\CO(m_t^2/\Lambda^2)$ of its critical value~\cite{Appelquist:1988as}. It was
shown in Ref.~\cite{Chivukula:1990bc} that ETC then generates a composite
complex EW doublet consisting of a scalar $H$ with vacuum expectation value
(vev)~$v = \CO(m_t)$ plus three Goldstone bosons, the longitudinal components
of $W$ and $Z$. The scalar has a large Yukawa coupling
$\Gamma_t \simeq m_t/v$ to the top quark and a mass
$M_H = \CO(m_t) \ll \Lambda$. Such a model might account for the Higgs
boson's exceptional lightness.

In Ref.~\cite{Lane:2014vca} we presented an explicit realization of such a
model. The model is a generalization of the topcolor model of Bardeen, Hill
and Lindner (BHL)~\cite{Bardeen:1989ds}, and followed that paper's
demonstration of the massless poles in fermion-antifermion scattering
channels that couple to the weak currents and of the massive pole in the
$0^+$ channel. An important difference with Refs.~\cite{Appelquist:1988as,
  Chivukula:1990bc, Bardeen:1989ds} is that three four-fermion interactions,
not just one, are required and there is a special relation among their
coupling strengths. The details are summarized below. Another point of
departure of the new model is that it pointed out the likely presence of
high-mass, $\rho$-like diboson resonances as an important, perhaps the most
important, experimental consequence of our model (see Secs.~3 and~4).

This model's approach differs from the popular view that the Higgs is a
pseudo-Goldstone boson; see, e.g., Ref~\cite{Contino:2010rs,
  Bellazzini:2014yua,Panico:2015jxa}. Not only is our model's $H$ not a PGB,
but there are no partners of the top quark and weak bosons to cancel their
quadratically divergent contributions to its mass. Rather, this quadratic
divergence is removed by the condition that $m_t$ and $m_T$ are much less
than $\Lambda$. This is a significant fine-tuning but, as explained below
Eq.~(\ref{eq:gapb}), this is the only one in the model. Thus, there is no
need to fine-tune partners' masses and couplings to explain why they haven't
been seen in LHC experiments.

In Ref.~\cite{Lane:2014vca} TC dynamics were not included. But TC cannot be
ignored. First, there must be an unbroken TC subgroup of ETC. If all its
symmetries were spontaneously broken, ETC would be infrared free at energies
below the ETC boson masses. It is unclear whether such a theory can be free
in the ultraviolet~\cite{Cheng:1973nv}. Second, at the scale
$\Ltc \simle 1\,\tev \ll \Letc = \Lambda$, the TC gauge coupling $\atc$
becomes strong enough that it can break EW symmetry all by itself. So, this
is a situation with two very different but nonetheless important energy
scales. ETC is the dominant force in driving EWSB and making the Higgs boson
light. But what sets the mass scale for the technihadrons, the bound states
of technifermions? Are they bound by TC alone or, like the Higgs boson, by
ETC, or by some cooperative combination? A major of purpose of this paper is
to include the effects of TC on EWSB and to estimate the mass scale of the
spin-one technihadrons. Because of their potential experimental importance,
we need to know whether they are much heavier than $H$ and, if so, whether
they are within reach of the LHC experiments.

That TC {\em must} play a minor role compared to ETC in EW symmetry breaking
was not emphasized in~Ref.~\cite{Lane:2014vca}, even though it was one of the
two main approximations of that paper. {\em The relative contributions that
  TC and ETC make in binding the spin-one technihadrons and generating their
  masses, and the requirement that the Higgs boson is much lighter than
  they, is what brings this issue to the fore.}

We now review the main results of Ref.~\cite{Lane:2014vca}: the fermions and
their ETC interaction, the fine-tuned gap equations for the fermions' masses,
the Higgs mass, and the principal results of EWSB. Then we preview the rest
of this paper.

The model of Ref.~\cite{Lane:2014vca} involved the third-generation quarks
and a single doublet of technifermions transforming under $(SU(2)\otimes
U(1))_{EW}$, ordinary color $\suc$ and technicolor $\sutc$ as follows:
\bea\label{eq:qT}
&& q_L = \left(\begin{array}{c} t\\b\\\end{array}\right)_L
\in ({\bs 2},\tx{\frac{1}{6}},{\bs 3}, {\bs 1}), \qquad
 t_R \in ({\bs 1},\tx{\frac{2}{3}},{\bs 3}, {\bs 1}),\quad
 b_R \in ({\bs 1},-\tx{\frac{1}{3}},{\bs 3}, {\bs 1}),\nn \\ \\\nn
&& T_L = \left(\begin{array}{c} U\\D\\\end{array}\right)_L
\in ({\bs 2},0,{\bs 1}, {\bs d_{TC}}), \qquad
 U_R \in ({\bs 1},\tx{\frac{1}{2}},{\bs 1}, {\bs d_{TC}}),\quad
 D_R \in ({\bs 1},-\tx{\frac{1}{2}},{\bs 1}, {\bs d_{TC}}).\nn
\eea
Here, ${\bs d_{TC}}$ denotes the $d_{TC}$-dimensional TC representation of
the technifermions, not necessarily the fundamental representation of
dimension $\Ntc$. Light quarks and leptons and other technifermions were not
dealt with, but they may be included, e.g., as outlined in
Ref.~\cite{Bardeen:1989ds}.

The ETC interaction inducing EWSB at energies below $\Lambda$ was taken to be
the straightforward generalization for these fermions of the
$SU(2)_L \otimes U(1)_R$-invariant model of Ref.~\cite{Bardeen:1989ds}:
\be\label{eq:LqT}
\CL_{ETC} = G_1\, \bar q^{ia}_L \,t_{Ra} \,\, \bar t^b_R \, q_{Lib}
          + G_2\, \left(\bar q^{ia}_L\, t_{Ra} \,\, \bar U^\alpha_R\,
            T_{Li\alpha} + {\rm h.c.} \right)
          + G_3\, \bar T^{i\alpha}_L\, U_{R\alpha} \,\, \bar U^\beta_R\,
          T_{Li\beta}  .
\ee
The $SU(2)_{EW}$ and color-$SU(3)_C$ and $\sutc$ indices, $i$ and $a,b$ and
$\alpha,\beta$ are summed over. This interaction is obtained by Fierzing ETC
contact terms of left times right-handed currents. The color and TC indices
appearing here do {\em not} correspond to exchange of massless color and TC
gluons. The couplings $G_{1,2,3}$ are positive and of
$\CO(1/\Lambda^2)$.\footnote{The $D$-technifermion gets no hard mass from ETC
  in this model. It is not difficult to add terms that generate $m_D \neq 0$,
  but not so easy to maintain both $m_D \cong m_U$ and $m_b \ll m_t$ at scale
  $\Lambda$. It was pointed out in Ref.~\cite{Chivukula:1990bc} that the
  renormalization group equations for the $U$ and $D$ Yukawa couplings have
  an infrared fixed point that tends to equalize $m_U$ and $m_D$ at $\Ltc$.}

In the neglect of EW interactions, the model has an
$(SU(2)_L\otimes U(1)_R)_q \otimes (SU(2)_L\otimes U(1)_R)_T$ flavor symmetry
that is explicitly broken to $SU(2)_L \otimes U(1)$ by the $G_2$-term. If
$\CL_{ETC}$ generates {\em both} $t$ and $U$ masses and $G_2 \neq 0$, this
flavor symmetry is spontaneously broken to $U(1)$ and just three Goldstone
bosons appear. In fact, $G_2$ must not equal zero; if it were, this would be
a two-Higgs doublet model with an extra triplet of Goldstone
bosons. They would acquire only very small EW masses~\cite{Eichten:1979ah}
and, so, are excluded experimentally. With $G_2 \neq 0$, this model has
exactly one Higgs boson. Its vev is $v = 246\,\gev$, setting the scale for
$m_{t,U}$; see Eqs.~(\ref{eq:fW},\ref{eq:zip}) below. The low-energy theory
below the ETC scale~$\Lambda$ is the standard model with spontaneously broken
$SU(2) \otimes U(1)$ and massive EW gauge bosons, a dynamical Higgs boson and
its couplings to~$t$ and~$U$, and technihadrons. Below the technihadron
masses, $M_{TC}$, their effects on SM Higgs couplings are suppressed by
$1/M_{TC}^2$, in accord with all measurements so far.\footnote{This was
  emphasized in Ref.~\cite{Bardeen:1989ds} in which the authors presented the
  effective Lagrangian with the Higgs boson, and discussed its
  renormalization.}

The TC interaction was neglected in Ref.~\cite{Lane:2014vca}, and
calculations were carried out in the Nambu--Jona-Lasinio (NJL) approximation
of large $d_{TC}$ and $N_C$. The gap equations for the hard masses $m_t$ and
$m_U$, assumed to be much less than $\Lambda$ and renormalized at the scale
$\Lambda$, are
\bea
\label{eq:gapt}
m_t &=& -\thalf G_1 \langle\bar tt\rangle -
\thalf G_2 \langle\bar UU\rangle \nn\\
    &=&  \frac{G_1 N_C m_t}{8\pi^2}\left(\Lambda^2 -
      m_t^2\ln\frac{\Lambda^2}{m_t^2}\right) 
          +\frac{G_2 d_{TC} m_U}{8\pi^2}\left(\Lambda^2 -
      m_U^2\ln\frac{\Lambda^2}{m_U^2}\right) \,;\\\nn\\
\label{eq:gapU}
m_U &=& -\thalf G_2 \langle\bar tt\rangle -
\thalf G_3 \langle\bar UU\rangle \nn\\
    &=&  \frac{G_2 N_C m_t}{8\pi^2}\left(\Lambda^2 -
      m_t^2\ln\frac{\Lambda^2}{m_t^2}\right) 
          +\frac{G_3 d_{TC} m_U}{8\pi^2}\left(\Lambda^2 -
      m_U^2\ln\frac{\Lambda^2}{m_U^2}\right) .
\eea
%
%
We can treat the dimensionalities $N_C$ and $d_{TC}$ as independent and even
continuous parameters for which the gap equations hold. Then, multiplying
Eq.~({\ref{eq:gapt}) by $m_U$ and Eq.~(\ref{eq:gapU}) by $m_t$, and varying
  $N_C$ and $d_{TC}$ independently, the two resulting equations can be true
  for nonzero $m_t$ and $m_U$ if and only if
\be\label{eq:refseq}
 \frac{G_1 N_C m_U}{8\pi^2} = \frac{G_2 N_C m_t}{8\pi^2}\,\,\,{\rm and}
\,\,\, \frac{G_2 d_{TC} m_U}{8\pi^2} = \frac{G_3 d_{TC} m_t}{8\pi^2}.
\ee
Hence, 
 \be\label{eq:Gequal}
 G_2 = G_1\frac{m_U}{m_t} = G_3\frac{m_t}{m_U}.
 \ee
Then, Eqs.~(\ref{eq:gapt}--\ref{eq:Gequal}) yield the condition:
\bea\label{eq:gapb}
&&\frac{G_1 N_C}{8\pi^2}\biggl(\Lambda^2 -
      m_t^2\ln\frac{\Lambda^2}{m_t^2}\biggr) 
          +\frac{G_3 d_{TC}}{8\pi^2}\biggl(\Lambda^2 -
      m_U^2\ln\frac{\Lambda^2}{m_U^2}\biggr) \nn \\
&&\quad = G_2 \biggl[\frac{N_C m_t}{8\pi^2 m_U}\biggl(\Lambda^2 -
      m_t^2\ln\frac{\Lambda^2}{m_t^2}\biggr)
      + \frac{d_{TC} m_U}{8\pi^2 m_t}\biggl(\Lambda^2 -
      m_U^2\ln\frac{\Lambda^2}{m_U^2}\biggr)\biggr] = 1.
\eea
It was shown in Ref.~\cite{Lane:2014vca} that $m_t$ and $m_U$ are comparable
and, so, the three $G_i$ are comparable as well.

Eq.~(\ref{eq:gapb}) is the expression of {\em strong} ETC in our model. This
raises the question of whether TC, the unbroken subgroup of ETC, must be
strongly coupled at the ETC scale. If it is, that would contradict the thesis
of this paper that TC is and must be a weak perturbation on ETC insofar as
triggering EWSB is concerned. But the $G_i$ are in fact {\em independent} of
the ETC gauge coupling, $g_{ETC}$. Just as in the standard weak interaction
at low energies, they are essentially equal to $g^2_{ETC}/M^2_{ETC}$, where
$M_{ETC} \propto g_{ETC}$ times a Goldstone boson decay constant. Thus, the
gauge coupling may be relatively weak while the four-fermion couplings are
strong in the sense of Eq.~(\ref{eq:gapb}).

Requiring $m_t, m_U \ll \Lambda$ is this model's {\em only} fine tuning. Once
Eq.~(\ref{eq:gapb}) is enforced in the fermion-antifermion scattering
amplitudes in the spin-zero channels, all other sensitivity to the cutoff
$\Lambda$ is logarithmic. The mass parameters $m_t$, $m_U$, $M_W$, $M_H$ and
$\Lambda$ are not independent. In the large-$N$ approximation, their
magnitude is set by requiring Eq.~(\ref{eq:zip}) below, and the Higgs mass
$M_H$ is then determined by $m_t$, $m_U$ and $N_C$, $d_{TC}$.

The fermion-antifermion scattering amplitudes (involving $t$ and/or $U$) have
a pole in the $0^+$ channel at squared c.m. energy $p^2 = M_H^2$, where $M_H$
is the solution of
\bea\label{eq:Hmass}
&& N_C m_t^2 (M_H^2-4m_t^2)\, 
\int_0^1 dx\, \ln\left(\frac{\Lambda^2}{m_t^2 - M_H^2 x(1-x)}\right)\nn\\
&& + d_{TC} m_U^2 (M_H^2-4m_U^2)
\int_0^1 dx\, \ln\left(\frac{\Lambda^2}{m_U^2 - M_H^2 x(1-x)}\right)  = 0.
\eea
This $M_H$ is the Higgs boson mass at scale $\Lambda$. A good approximation
to the solution of Eq.~(\ref{eq:Hmass}) is
\be\label{eq:Hmassb}
M_H = 2\sqrt{\frac{N_C m_t^4 + d_{TC} m_U^4}{N_C m_t^2 + d_{TC} m_U^2}}.
\ee
Thus, $M_H$ is indeed of order $m_t,m_U$ and all these masses are much less
than $\Lambda$ because the ETC couplings have been tuned to be very close to
the critical point at which EWSB first occurs.

The fermion-antifermion scattering amplitudes in the charged and neutral
pseudoscalar channels have Goldstone poles at $p^2 = 0$. These poles appear
in the $W$ and $Z$ propagators, $g_2^{-2}D_W(p)$ and $(g_1^2 +
g_2^2)^{-1}D_Z(p)$, with residues
\be\label{eq:fW}
f_W^2(p^2) = \frac{1}{16\pi^2}\int_0^1 dx\, x
    \biggl[N_C m_t^2 \ln\biggl(\frac{\Lambda^2}{m_t^2 x - p^2 x(1-x)}\biggr)
+ d_{TC} m_U^2 \ln\biggl(\frac{\Lambda^2}{m_U^2 x - p^2
  x(1-x)}\biggr)\biggr]\,
\ee
and
\bea\label{eq:fZ}
f_Z^2(p^2) &=& \frac{1}{32\pi^2}\int_0^1 dx\,
    \biggl[N_C\,m_t^2 \ln\biggl(\frac{\Lambda^2}{m_t^2 - p^2 x(1-x)}\biggr)
 + d_{TC}\,m_U^2 \ln\biggl(\frac{\Lambda^2}{m_U^2 - p^2
   x(1-x)}\biggr)\biggr]\nn\\
&+& \frac{N_C p^2}{16\pi^2}\int_0^1dx\, {\tthird} x(1-x)
\ln\biggl(\frac{-p^2 x(1-x)}{m_t^2 - p^2 x(1-x)}\biggr).
\eea
The EW mass scale is introduced by setting
\be\label{eq:zip}
f_W^2(0) = 1/(4\sqrt{2}G_F) = (123\,\gev)^2\ \cong M_W^2/g_{2W}^2(0).
\ee
The $\rho$-parameter,
\be\label{eq:rho}
\rho \cong \frac{f_W^2(0)}{f_Z^2(0)} =
\frac{\left[N_C\,m_t^2\left(\ln(\Lambda^2/m_t^2)+\thalf\right) +
d_{TC} \,m_U^2\left(\ln(\Lambda^2/m_U^2)+\thalf\right)\right]}
{\left[N_C \,m_t^2 \ln(\Lambda^2/m_t^2) +
d_{TC} \,m_U^2 \ln(\Lambda^2/m_U^2)\right]} ,
\ee
is just a few percent greater than one. This is spurious. The deviation of
$\rho$ from unity in Eq.~(\ref{eq:rho}) is due to the factors of 1/2 in the
numerator. Those factors should not have been included in
Ref.~\cite{Lane:2014vca} because the calculations of $f_W^2(0)$ and
$f_Z^2(0)$ where done in the leading-log approximation (in the cutoff/ETC
scale $\Lambda$). Corrections to this approximation are unknown.

Table~1 contains numerical results for the model obtained from a simple
scheme described in Ref.~\cite{Lane:2014vca}.
 \begin{table}[!ht]
     \begin{center}{
  \begin{tabular}{|c|c|c|c|c|c|}
  \hline
 $\Lambda$ & $m_t$ & $m_U$ & $M_H$ &$\Gamma_t$ & $v=\sqrt{2} m_t/\Gamma_t$ \\
  \hline\hline
 $20\,\tev$ & $134\,\gev$ & $167\,\gev$ & $330\,\gev$ & 0.783  & $242\,\gev$ \\
 $500\,\tev$ & $118\,\gev$ & $126\,\gev$ & $250\,\gev$ & 0.685  & $244\,\gev$ \\
  \hline\hline
 $\Lambda$ & $\rho$ & $g_1$ & $g_2$ & $M_W$({\rm pole}) &$M_Z({\rm pole})$ \\
   \hline\hline
 $20\,\tev$  & 1.0520 &0.3941  & 0.7187 & 80.8  & 91.0  \\
 $500\,\tev$ & 1.0301 &0.4230  & 0.7714 & 80.6  & 93.0 \\
\hline
 \end{tabular}}
\caption{The fermion masses, Higgs boson mass, $\rho$-parameter,
  $(SU(2) \otimes U(1))_{EW}$ couplings and the $W$, $Z$-pole masses
  calculated for ETC scales $\Lambda = 20$ and $500\,\tev$. The top mass is
  an input determined by renormalizing from its value of $173\,\gev$. The
  Higgs boson's vev~$v = \sqrt{2}m_t/\Gamma_t$ is determined as a check on
  the calculation of the $\bar tt$ scattering amplitude in the scalar
  channel, where $\Gamma_t^2(\Lambda)/2$ is the residue of the Higgs pole.
  The calculation scheme used is described in Ref.~\cite{Lane:2014vca}. As
  noted in the text, the deviations of $\rho$ from one are not reliably
  calculated.\label{tab:numerical}}
 \end{center}
 \end{table}

 In Sec.~2 we discuss the difficulty of adding an interaction involving
 dynamical TC-gluon exchange to the ETC contact interaction in
 Eq.~(\ref{eq:LqT}) and propose an approximation that surmounts the problem
 for fermion-antifermion scattering in the spin-zero channel. The
 approximation is inspired by analyses of the effect of TC on the
 Schwinger-Dyson equation for the technifermion dynamical mass function,
 $\Sigma(p)$\cite{Appelquist:1988fm, Takeuchi:1989qa}. In Sec.~3 we take up
 the matter of estimating the masses of the lightest spin-one vector and
 axial vector bound states, analogs of $\rho$, $\omega$ and~$a_1$. We shall
 refer to them as $\rho_H$, $\omega_H$ and $a_H$ to emphasize their relation
 to the composite Higgs boson $H$. In this strong-ETC model, it is not
 obvious {\em a priori} whether their masses are of order $\Letc$, $\Ltc$ or
 something else, though that is a question of obvious phenomenological
 importance. We present a calculation that suggests they are of $\CO(\Ltc)$.
 As in any strong interaction theory, a more precise estimate is technically
 difficult. Assuming they are within reach of LHC Runs~$2+3$, their LHC
 phenomenology is discussed in Sec.~4.\footnote{A preliminary discussion
   appeared in Ref.~\cite{Lane:2014vca}.} There we review our recent
 proposal~\cite{Lane:2015fza} that $\rho_H$ and $a_H$ are the source of the
 apparent diboson ($VV$ and $VH$, where $V=W,Z$) resonances near $2\,\tev$
 observed by ATLAS and CMS in their Run~1
 data~\cite{Aad:2015owa,Aad:2015ufa,Aad:2015yza,Khachatryan:2014gha,
   Khachatryan:2014hpa, Khachatryan:2016yji} and we propose refined tests
 of our hypothesis for Runs~$2+3$. New limits on diboson resonances from
 Run~2 data of $36\,\ifb$ are also discussed in Sec.~4.

 There has been much previous work using the NJL
 mechanism~\cite{Nambu:1961tp,Nambu:1961fr} to describe the Higgs boson,
 including especially Refs.~\cite{Bardeen:1989ds,Nambu:1989jt,
   Miransky:1988xi,Miransky:1989ds}. Topcolor led to the top-seesaw models of
 Dobrescu and Hill~\cite{Dobrescu:1997nm} and Chivukula, et
 al.~\cite{Chivukula:1998wd} and, more recently, Refs.~\cite{Fukano:2012qx,
   Fukano:2013kia}.\footnote{The last two papers contain a large bibliography
   of related work.} Bar-Shalom and collaborators proposed a ``hybrid model''
 with a dynamical Higgs-like scalar plus an elementary scalar to describe
 $H$~\cite{Bar-Shalom:2013hda, Geller:2013dla}. They used an NJL Lagrangian
 with fourth generation quarks interacting via a topcolor interaction with
 scale $\Lambda \sim 1\,\tev$ to generate the dynamical scalar. Apart from
 the use of the NJL bubble approximation, these models do not resemble ours,
 and the use of fourth generation quarks is reminiscent of the top-seesaw
 mechanism. The top-seesaw models involve mixing the top quark with another
 quark which is a weak isosinglet. That is not what happens in our model. The
 technifermion $U$ carries technicolor, not ordinary color, and its
 left-handed component is in a weak isodoublet, like the quarks and
 leptons. It does not mix with the top or any other quark.

 Di~Chiara, {\em et al.}, proposed a model of $H$ based on TC and
 ETC~\cite{DiChiara:2014gsa,DiChiara:2014uwa}, using an ETC Lagrangian
 similar to Eq.~(\ref{eq:LqT}). Their model bears no further resemblance to
 ours. They assume that ETC plays no role in EWSB. But, through a sequence of
 calculations, they argue that ETC lowers their Higgs boson's mass from
 $\CO(1\,\tev)$ to $125\,\gev$. Finally, the authors of
 Ref.~\cite{vonGersdorff:2015fta} proposed an interesting variation on the
 Higgs boson as a PGB of the familiar $SO(5) \to SO(4)$ model. They used
 strong ETC-like contact interactions to drive this symmetry breakdown, and
 constructed a UV completion of this model.

\begin{figure}[!t]
 \begin{center}
\includegraphics[width=6in, height=2in]{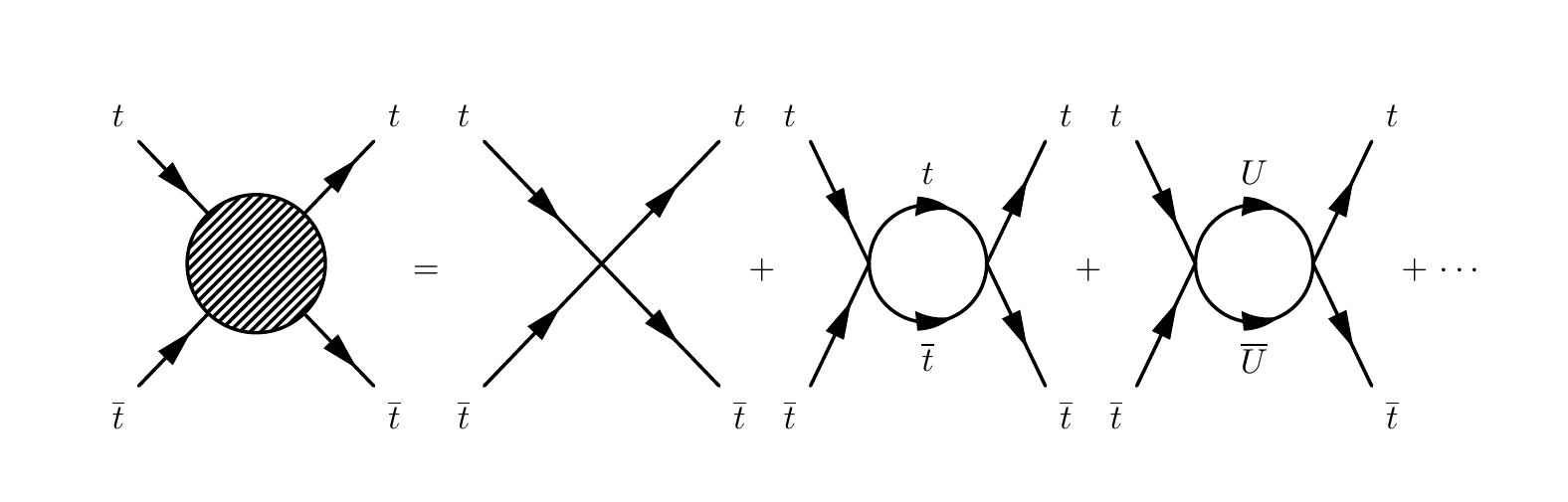}
\caption{The $\bar tt$ scattering amplitude in the $J^P=0^+$ channel with the 
  four-fermion kernel given by the terms in Eq.~(\ref{eq:Lzeroplus}).}
  \label{fig:TCETC_H_Fig1}
 \end{center}
 \end{figure}

\section{Adding TC to Strong ETC}

In Ref.~\cite{Lane:2014vca}, the Higgs and Goldstone bosons were seen as
poles in the fermion-antifermion scattering amplitudes calculated in the
large-$N$, weak-TC limit. Figure~\ref{fig:TCETC_H_Fig1} shows the first few
terms in $\bar tt \to \bar tt$ in the $J^P = 0^+$ channel. The four-fermion
vertices are the appropriate terms in
\bea\label{eq:Lzeroplus}
\CL_{ETC} = \tfourth\left[G_1\, \bar t^{a} \,t_{a} \,\, \bar t^b \, t_{b}
          + G_2\, \left(\bar t^{a}\, t_{a} \,\, \bar U^\alpha\,
            U_{\alpha} + {\rm h.c.} \right)
          + G_3\, \bar U^{\alpha}\, U_{\alpha} \,\, \bar U^\beta\,
          U_{\beta}\right]  .
\eea
From Eq.~(\ref{eq:Gequal}), $G_2^2 = G_1 G_3$, and this condition makes the
scattering amplitudes geometric sums, with poles corresponding to the Higgs
and three Goldstone bosons. The Higgs pole-mass condition
Eq.~(\ref{eq:Hmass}) follows once Eq.~(\ref{eq:gapb}) is imposed to eliminate
the $\Lambda^2$-divergence in the $0^+$ scattering amplitude.

In the large-$N$ approximation, the inclusion of TC-gluon exchange between
the technifermions is accomplished by using the kernel $\CK_{0^+}$ in
Fig.~\ref{fig:TCETC_H_Fig2}. The TC-gluon term of this kernel is the familiar
ladder approximation. The difficulty with it is how to deal with the momentum
carried by the TC-gluon and, worse, whether the sum is a geometric series for
which something like Eq.~(\ref{eq:gapb}) eliminates the
$\Lambda^2$-divergence.

\begin{figure}[!t]
 \begin{center}
\includegraphics[width=6in, height=2in]{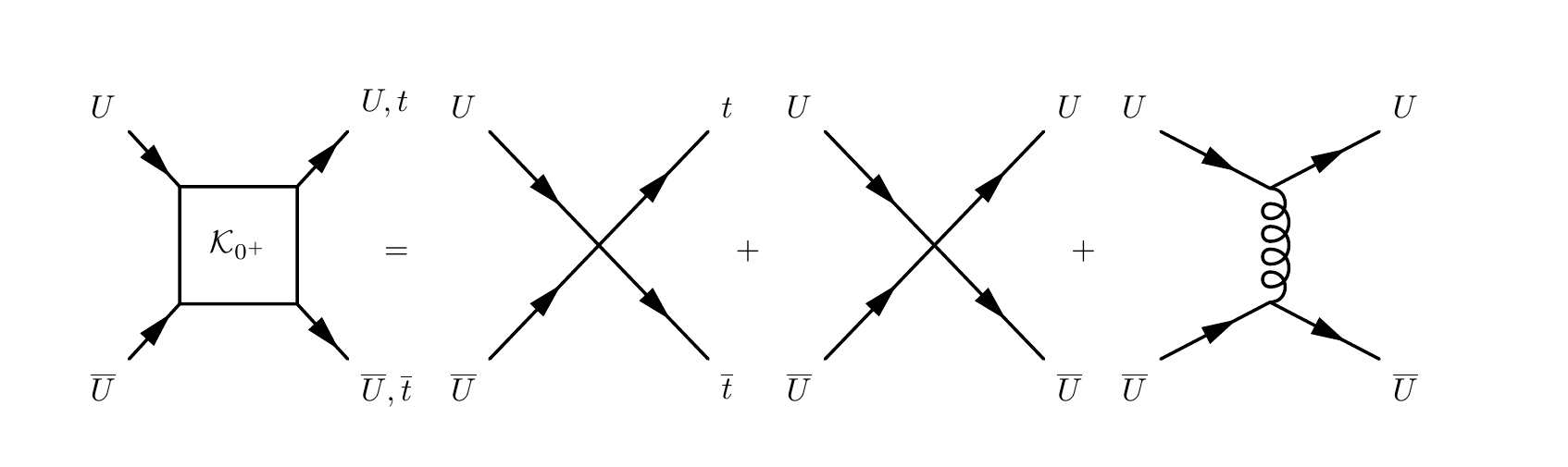}
\caption{The kernel for scattering of $\bar UU \to \bar tt$ and $\bar UU
  \to \bar UU$ including ETC contact terms and one-TC-gluon exchange.}
  \label{fig:TCETC_H_Fig2}
 \end{center}
 \end{figure}

 The only situation we know in which the ETC$+$TC kernel in
 Fig.~\ref{fig:TCETC_H_Fig2} has been used successfully is in studies of the
 dynamical mass function $\Sigma(p^2)$ in the technifermion propagator
 $S^{-1}(p) = \slashchar{p} A(p^2) - \Sigma(p^2)$ (where $A(p^2) = 1$ in the
 Landau gauge ladder approximation)~\cite{Appelquist:1988fm,
   Takeuchi:1989qa}. Remembering that the ETC boson mass $\Lambda$ is a
 physical cutoff of momentum integrals whose integrands are strongly damped
 above $\Lambda$, a good approximation to the Schwinger-Dyson gap equation
 for $\Sigma(p^2)$ is (for zero bare mass and Euclidean momentum
 $p \simle \Lambda$)
\be\label{eq:SDgap}
\Sigma(p^2) = \lambda \int_0^{\Lambda^2} dk^2\, \frac{k^2}{\Lambda^2}
\frac{\Sigma(k^2)}{k^2 + \Sigma^2(k^2)} 
   +\frac{1}{4\alpha_c}\int_0^{\Lambda^2} dk^2\, \atc(M^2)\frac{k^2}{M^2}
\frac{\Sigma(k^2)}{k^2 + \Sigma^2(k^2)}.
\ee
We consider a simplified model with just $G_3$ contributing to $\Sigma$. Then
$\lambda = G_3 d_{TC}\Lambda^2/8\pi^2$; $\atc$ is the running TC gauge
coupling; $\alpha_c$ is the critical value of $\atc$ for spontaneous chiral
symmetry breaking in a pure-technicolor theory~\cite{Cohen:1988sq}; its value
in the ladder approximation is $\pi/3C_2({\bf d}_{TC})$; finally,
$M^2 = {\rm max}(k^2,p^2)$.

In a pure ETC theory, $\Sigma(0) = 0$ for $\lambda < 1$, there is a
(presumed) second-order phase transition at $\lambda = 1$, and $\Sigma(0)$
rises rapidly to $\CO(\Lambda)$ just above the transition. In a pure
asymptotically-free TC theory, $\atc$ reaches $\alpha_c$ at a scale
$\Lambda_c$, $\Sigma(0) \simeq \Sigma(\Lambda_c) = \CO(\Lambda_c)$, and
$\Sigma(p^2)$ falls off approximately as $\Lambda_c^3/p^2$ when $\atc$
becomes weak~\cite{Lane:1974he}. Ref.~\cite{Appelquist:1988fm} studied
Eq.~(\ref{eq:SDgap}) for {\em constant} $\atc$. For $\atc < \alpha_c$, the
behavior of $\Sigma(0)$ was as in a pure-ETC theory except that, for
$\atc < \alpha_c$, the phase transition occurred at
\be\label{eq:lama}
\lambda_{\atc} = \left[\frac{1+\sqrt{1-\atc/\alpha_c}}{2}\right]^2.
\ee

Takeuchi studied the gap equation for a running~$\atc$ governed by the
one-loop beta function $\beta(\atc) = -b_1\atc^2$, with
$b_1 > 0$~\cite{Takeuchi:1989qa}. So long as $\Lambda_c \ll \Lambda$ (as we
expect), he found that $\Sigma(0) = \CO(\Lambda_c)$ for
$\lambda < \lambda_{\atc}$.\footnote{Takeuchi's definition of $\Lambda_c$ is
  lower than the scale at which $\atc = \alpha_c$. This appears to be an
  artifact of his calculation procedure at low momenta.} Here,
$\atc \simeq \atc(\Lambda)$. At this critical value of $\lambda$, there is a
smooth but rapid transition up to $\Sigma(0) = \Lambda/{\rm few}$. The
transition is more abrupt for small $\atc(\Lambda)/\alpha_c$ so that
$\lambda_{\atc} \sim 1$. The reason that $\atc(\Lambda)$ is the controlling
coupling for $\lambda_{\atc}$ is that, for this $\beta$-function and
$\Lambda_c \ll \Lambda$, $\atc \simeq \atc(\Lambda) \ll \alpha_c$ and it is
slowly running for most of the momentum range in the gap equation
integral. We have verified Takeuchi's results for a more realistic walking-TC
$\beta$-function, one with an infrared fixed point~\cite{Lane:1991qh}. We
also studied the momentum dependence of $\Sigma(p^2)$. For
$\lambda < \lambda_{\atc(\Lambda)}$, we found that $\Sigma$ is small and
falls off approximately as $1/p^2$ for $\Lambda_c \simle p \simle \Lambda$,
as for a pure-TC dynamical mass. At the critical $\lambda$, $\Sigma(p)$ rises
rapidly to $\CO(\Lambda/10)$ and then remains nearly constant in $p$, as for
a hard mass.

This is an important result for us. In the weak dynamical-TC case needed for
a light composite Higgs with $M_H^2 \ll M_{\rho_H}^2$, this behavior of
$\Sigma$ is nearly what we get in the complete neglect of TC: it is much
smaller than $\Lambda$ below $\lambda_{\atc}$ and rises {\em abruptly} above,
almost to $\CO(\Lambda)$. (Had $\atc(\Lambda)$ been large, the transition
from small to large $\Sigma$ would have been gradual and there could be no
large separation between $H$ and $\rho_H$~masses.) The critical
$\lambda_{\atc}$ is smaller than one because, to a good approximation, TC
produces an interaction {\em in the spin-zero channels} of the same form and
sign as the $G_3$-term in $\CL_{ETC}$. Thus, a smaller value of $G_3$, i.e.,
$\lambda$, is needed to trigger the phase transition. The critical value of
the sum of the two interaction strengths is still fixed by a condition like
Eq.~(\ref{eq:gapb}); i.e., the effective $G_3$ in $\CL_{ETC}$ is essentially
unchanged. Since $\CL_{ETC}$ is the interaction determining the Higgs and
Goldstone poles and their couplings to fermions in the large-$N$ limit, the
results reviewed in Sec.~1 are also unchanged.

To see this in detail, we use the fact that the TC coupling involved in the
EW phase transition is approximately $\atc(\Lambda)$. The relevant TC
interaction then involves exchange of a technigluon with {\em Euclidean}
momentum transfer $\approx -\Lambda^2$,
\be\label{eq:LTC}
\CL_{TC} = -\frac{3\pi\atc(\Lambda)}{2\Lambda^2} \sum_A \bar T\gamma^\mu t_A T
\, \bar T\gamma_\mu t_A T,
\ee
where $T = (U,D)$ is the technifermion doublet, $t_A$ are the TC generators
in the representation ${\bs d_{TC}}$, and other indices are suppressed. A
factor of 3/4 has been introduced into $\CL_{TC}$ to compensate for using
Landau instead of Feynman gauge.\footnote{Although this term is
  isospin-symmetric, its strength is not sufficient to produce
  $m_D \neq 0$.} Use
\be\label{eq:tAsum}
\sum_A (t_A)_\alpha^{\beta} (t_A)_{\gamma}^{\delta} = 
\frac{C_2({\bf d}_{TC}) d_{TC}}{d_{TC}^2-1} \left(\delta_{\alpha}^{\delta}
  \delta_{\gamma}^{\beta} - \frac{1}{d_{TC}}\delta_{\alpha}^{\beta}
  \delta_{\gamma}^{\delta}\right)
.
\ee
Then, in the large-$\Ntc$ limit, $\CL_{TC}$ Fierz-transforms into
\bea\label{eq:LTCF}
\CL_{TC} &=& \frac{3\pi C_2({\bf d}_{TC})\atc(\Lambda)}{d_{TC}\Lambda^2}
\biggl[\bar T^{i\alpha} T_{j\alpha} \,\bar T^{j\beta} T_{i\beta}
- T^{i\alpha} \gamma_5 T_{j\alpha} \,\bar T^{j\beta} \gamma_5 T_{i\beta}\nn\\
&\quad& -\thalf \bar T^{i\alpha} \gamma^\mu T_{j\alpha} \,\bar T^{j\beta}
\gamma_\mu T_{i\beta}
- \thalf \bar T^{i\alpha} \gamma^\mu \gamma_5 T_{j\alpha} \,\bar T^{j\beta}
\gamma_\mu \gamma_5 T_{i\beta}\biggr].
\eea
Adding this to the $G_3$ term in Eq.~(\ref{eq:Lzeroplus}), the effective
$\lambda$ is
\be\label{eq:lambdaeff}
\lambda_{eff} = \frac{G_3 d_{TC} \Lambda^2}{8\pi^2} +
\frac{3 C_2({\bf d}_{TC})\atc(\Lambda)}{4\pi} = \lambda +
\frac{\alpha_{TC}(\Lambda)}{4\alpha_c} .
\ee
For the critical value $\lambda_{eff} = 1$,
$\lambda = 1 - \atc(\Lambda)/4\alpha_c$. This is less than 20\% higher than
$\lambda_{\atc(\Lambda)}$ for $\alpha_{TC}(\Lambda)/\alpha_c < 0.5$, which is
the range that Takeuchi considered. Thus, our approximation for $\CL_{TC}$
captures well the main effect of adding TC to ETC in the gap equation and
spin-zero scattering amplitudes.

We address at this point the following question: Is there an additional
spontaneous breaking of EW (or any other) symmetry when TC becomes strong and
forms the condensate $\langle \bar DD \rangle$? The answer is no. The EW
symmetry is already broken to $U(1)_{EM}$ by the ETC
interaction. Furthermore, there is no appreciable contribution to the EW
order parameter~$v$ because $D$-condensate gives rise to no Goldstone
boson. The chiral current $\bar D \gamma_\mu \gamma_5 D$ has a TC-anomalous
divergence and explicit breaking of this symmetry is $\CO(\Ltc)$.

\section{Masses of the Spin-One Technihadrons}

As we stressed at the outset, the challenge for a TC-based composite Higgs
model is to explain convincingly why $H$ is much lighter than the
lowest-lying spin-one technihadrons. In our model, there is the additional
matter that there are two scales, $\Ltc$ and $\Letc = \Lambda$. Which of
these controls $M_{\rho_H}$?  If it is just $\Lambda$, are these masses of
that order or, as for the Higgs, very much lighter? In this section we
present an argument suggesting they are at least as heavy as $\Ltc$ and
therefore well above the Higgs mass. For this, we {\em assume} that
$M_{\rho_H},\dots$ are due entirely to an ETC interaction and find that this
results in unphysical or implausible masses for these states.

We start by considering a simplified model with the doublet $T=(U,D)$ as the
only fermions. Its $SU(2)_L \otimes U(1)_R$ invariant ETC interaction is
\be\label{eq:LTone}
\CL_T = G_3 \bar T_L^{i\alpha} U_{R\alpha} \bar U_R^\beta T_{Li\beta}.
\ee
This interaction produces nonzero $m_U$, but not $m_D$, if
\be\label{eq:gapUU}
\frac{G_3 d_{TC}}{8\pi^2}\left(\Lambda^2 -
  m_U^2\ln\frac{\Lambda^2}{m_U^2}\right) = 1.
\ee
While $\CL_T$ can generate a light Higgs boson and three Goldstone bosons,
it has the wrong chiral structure to generate masses for the spin-one
technihadrons in the large-$\Ntc$ limit. Therefore, we expand it to include
terms capable of this. We assume that ETC generates $VV$ and $AA$ contact
interactions which add to $\CL_T$. For simplicity, we can take them to be
$V$-$A$ symmetric and flavor-$U(2)$ invariant without affecting our argument:
\be\label{eq:LTtwo}
\CL_T = G_3 \biggl[\bar T_L^{i\alpha} U_{R\alpha}\, \bar U_R^\beta
  T_{Li\beta} 
  - \tfourth\delta\sum_{a=0}^3\biggl(\bar T^\alpha \gamma^\mu \tau_a T_\alpha \,
\bar T^\beta \gamma_\mu \tau_a T_\beta + (\gamma_\mu \to \gamma_\mu
\gamma_5)\biggr)\biggr],
\ee
where $\tau_a$ are Pauli matrices acting in the $(U,D)$-flavor space. The
parameter $\delta$ allows freedom in the choice of the ETC coupling of the
$VV$ and $AA$ terms. We write (with a unit $\rho_H$ coupling to the $U(2)$
current)
\be\label{eq:rhomat}
\lvac \bar T^\alpha \gamma_\mu \frac{\tau_A}{2} T_\alpha|\rho_B(p)\rangle =
\epsilon_\mu(p) \delta_{AB} ,
\ee
where $p^\mu\epsilon_\mu(p) = 0$. Then, to leading order in $\Ntc$, the
technivector masses are given by the poles in the $\rho_A \to \rho_B$
amplitude
\bea\label{eq:Trho}
\CT_{AB}(p) &=& \epsilon^{\mu *}(p)\epsilon^\nu(p)(-2\delta G_3) 
\biggl[g_{\mu\nu}\delta_{AB} - \thalf \delta G_3d_{TC} I^{AB}_{\mu\nu}(p) \nn\\
 &\quad& + (-\thalf \delta G_3d_{TC})^2 \sum_{C}
 I^{AC}_{\mu\lambda}(p)I^{CB}_{\lambda\nu}(p) + \cdots\biggr] \nn\\
&=& \epsilon^{\mu *}(p)\epsilon^\nu(p)(-2\delta G_3) 
\biggl[\biggl(1 + {\thalf} \delta G_3 d_{TC}
I(p)\biggr)^{-1}\biggr]_{\mu\nu}^{AB},
\eea
where
\be\label{eq:Imunu}
I^{AB}_{\mu\nu}(p) = i\int\frac{d^4k}{(2\pi)^4} {\rm Tr}
\left[\left(\frac{\slashchar{k}+\slashchar{p}+M}{(k+p)^2-M^2}\right)
  \gamma_\mu \tau_A
\left(\frac{\slashchar{k}+M}{k^2-M^2}\right) \gamma_\nu\tau_B\right].
\ee
In this model, with $m_D = 0$, the fermion mass matrix is
\be\label{eq:Tmass}
M = \left(\begin{array}{cc}
m_U & 0\\
0 & 0\\
\end{array}\right).
\ee
The momentum integral~(\ref{eq:Imunu}) is cutoff at $\Lambda$, just as the
ones in the spin-zero channels were, giving
\bea\label{eq:Iresults}
&&I_{\mu\nu}^{11,22}(p) = -\frac{\Lambda^2}{4\pi^2}g_{\mu\nu} + \frac{1}{\pi^2}
\int_0^1 dx\bigl[(p_\mu p_\nu - p^2 g_{\mu\nu})x(1-x) +{\thalf} m_U^2\, x
g_{\mu\nu}\bigr]\nn\\
&& \qquad\qquad\qquad \times \ln\biggl(\frac{\Lambda^2 + m_U^2 x - p^2
  x(1-x)}{ m_U^2 x - p^2 x(1-x)}\biggr)\,; \\
&& I_{\mu\nu}^{33,00}(p) = -\frac{\Lambda^2}{4\pi^2}g_{\mu\nu} + \frac{1}{2\pi^2}
\int_0^1 dx \,(p_\mu p_\nu - p^2 g_{\mu\nu})x(1-x) \nn\\
&&\qquad \qquad \qquad \times \biggl[\ln\biggl(\frac{\Lambda^2 + m_U^2 -
  p^2 x(1-x)}{m_U^2 - p^2 x(1-x)}\biggr) +  
\ln\biggl(\frac{\Lambda^2 - p^2 x(1-x)}{-p^2 x(1-x)}\biggr) \biggr]\,; \\
&& I_{\mu\nu}^{30,03}(p) = \frac{1}{2\pi^2}
\int_0^1 dx \, (p_\mu p_\nu - p^2 g_{\mu\nu})x(1-x)\nn\\
&& \qquad \qquad \qquad \times
\biggl[\ln\biggl(\frac{\Lambda^2 +m_U^2 - p^2 x(1-x)}{m_U^2 - p^2 x(1-x)}\biggr)
- \ln\biggl(\frac{\Lambda^2 - p^2 x(1-x)}{-p^2 x(1-x)}\biggr)\biggr] .
\eea
The $p_\mu p_\nu$ terms in these integrals do not contribute to $\CT_{AB}$.
Then, in the leading-log approximation, the $\rho_H$-$\omega_H$ mixing term
is negligible and the poles in $\CT_{AB}$ are at 
\bea\label{eq:IAA}
 1 - \frac{\delta G_3 d_{TC} \Lambda^2}{8\pi^2} 
- \frac{\delta G_3 d_{TC}(p^2 -  \textstyle{\frac{3}{2}}m_U^2)}{12\pi^2}
\ln\biggl(\frac{\Lambda^2}{m_U^2}\biggr) &=& 0  \qquad{\rm for}\,\,
A=B=1,2\,; \\ 
 1 -\frac{\delta G_3 d_{TC} \Lambda^2}{8\pi^2} -\frac{\delta G_3 d_{TC}
   p^2}{12\pi^2}
\ln\biggl(\frac{\Lambda^2}{m_U^2}\biggr) &=& 0 \qquad{\rm for}\,\, A=B=3,0. 
\eea
Using the gap Eq.~(\ref{eq:gapUU}), the poles in $\CT_{11,22}$ and
$\CT_{33,00}$ are at $\bar p^2$ satisfying
\bea\label{rhopoles}
\bar p^2\ln\biggl(\frac{\Lambda^2}{m_U^2}\biggr) &=&
\frac{3}{2}\biggl(\frac{1-\delta}{\delta}\biggr)
\biggl[\Lambda^2 - m_U^2\ln\biggl(\frac{\Lambda^2}{m_U^2}\biggr)\biggr]
\qquad{\rm for}\,\, \rho_H^\pm\,; \\ 
\bar p^2\ln\biggl(\frac{\Lambda^2}{m_U^2}\biggr) &=&
\frac{3}{2\delta}\biggl[(1-\delta)\Lambda^2 - 
m_U^2 \ln\biggl(\frac{\Lambda^2}{m_U^2}\biggr)\biggr] \qquad{\rm for}\,\,
\rho_H^0,\omega_H.
\eea
This is unphysical unless $0 < \delta < 1$ for $\rho_H^\pm$ and
$0 < \delta < 1-m_U^2/\Lambda^2 \ln(\Lambda^2/m_U^2)$ for $\rho_H^0$ and
$\omega_H$. For $\delta$ at its upper limit, $\bar p^2 \simeq 0$, i.e., very
much less than $\Lambda^2_{TC}$. We believe this is unreasonable because no
symmetry is responsible for such light masses. Our calculations break down
beyond the $\Lambda$-cutoff, so $M_{\rho_H} \simge \Lambda$ is an unreliable
result. Over a large part of the physical range of~$\delta$,
$\Lambda^2_{TC} \ll \bar p^2 \simeq \Lambda^2/\ln(\Lambda^2/m_U^2) <
\Lambda^2$.
We cannot exclude this. But, for a mass we have assumed is generated solely
by strong ETC, it seems implausible to us. A more believable result is that
TC generates the $\rho_H$ and $\omega_H$ masses and that they are of order
$\Ltc$, the scale at which $\atc$ becomes large and TC interactions
confine. In both the latter two cases, the $\rho_H$, $\omega_H$ masses are
significantly larger than the Higgs mass, and that is a necessary condition
for the viability of this type of model.

Let us extend this argument to the full $G_1$-$G_2$-$G_3$ model. There are
two obvious possibilities for the $VV+AA$ terms: we could add just the
$\delta G_3$ interaction as we did in Eq.~(\ref{eq:LTtwo}) or we could add
similar terms with the appropriate coefficient, $-\tfourth \delta G_i$, to
all three interactions. The gap-equation condition is now given by
Eq.~(\ref{eq:gapb}). In the first case, the poles are always at
$\bar p^2 \sim \Lambda^2/\ln(\Lambda^2/m_U^2)$, which we believe is
implausible. The second case is similar to the pure-$G_3$ model discussed
above. Finally, similar results and conclusions hold for ETC-generated masses
of the axial vectors $a_H$; they are either unreasonably small or much larger
than $\Ltc$ but smaller than $\Lambda$. The conclusion we draw is that an ETC
origin of the technivector masses is less plausible than that they arise from
the confined TC interactions and are of $\CO(1\,\tev)$.

\section{Phenomenology of $\rho_H$ and $a_H$}

Preliminary remarks about the phenomenology of the model's technifermion
bound states were made in Ref.~\cite{Lane:2014vca}. They included, in
particular, the expectations that: (1) the most accessible low-lying states,
in addition to the Higgs~$H$ and longitudinal weak bosons~$W_L$ and~$Z_L$
(which really are bound by the ETC interaction, Eq.~(\ref{eq:LqT})), are the
spin-one, techni-isospin one and zero $\rho$ and $\omega$-like composites;
(2) their masses are $\sim$ 1/2--$2\,\tev$ and they are produced at the LHC
via the Drell-Yan process; (3) their principal decay modes would be to
$W^+_L W^-_L, W^\pm_L, Z_L$ or $W^+_L W^-_L Z_L$ and to $W_L H$, $Z_L H$. In
this section we refine -- and correct -- these expectations, presenting some
specific predictions of production and decay rates. We concentrate on the
$I=1$ vectors and axial vectors, $\rho_H$ and $a_H$, which have simple
two-body decay modes and Drell-Yan-size production rates.

The technicolor interaction governing $\rho_H$ and $a_H$ is invariant under
parity and techni-isospin of $(U,D)$. This symmetry is broken by the
electroweak gauge interaction and the $U$--$D$ mass difference. The first is
$\CO(\alpha)$ and the second is an $I=1$ operator and does not contribute to
mass splitting within the isotriplet multiplets. Furthermore, because the ETC
interaction, Eq.~(\ref{eq:LqT}), is tuned to be close to the EW phase
transition, we expect that $\rho_H$ and $a_H$ are nearly parity-doubled
triplets with $M_{\rho_H} \cong M_{a_H}$. Thus, they can be adequately
described as the gauge bosons of a hidden local symmetry
(HLS)~\cite{Bando:1987br}, $SU(2)_L \otimes SU(2)_R$, with equal gauge
couplings $g_L = g_R \equiv g_{\rho_H}$~\cite{Casalbuoni:1995qt, Lane:2009ct,
  Bellazzini:2012tv,Appelquist:2015vdl}. This coupling is analogous to
$g_{\rho\pi\pi}$ and is expected to be large, $g_{\rho_H} \simeq 3$-5. The
equality $g_L = g_R$ makes the $\rho_H$-$a_H$ contribution to the
$S$-parameter~\cite{Kennedy:1988sn, Peskin:1990zt,Golden:1990ig,
  Holdom:1990tc, Altarelli:1991fk}
small~\cite{Casalbuoni:1995qt,Lane:2009ct}. The dimension-three and~four
interactions of $\rho_H,a_H$ with EW gauge bosons and the Higgs respect
parity-invariance up to EW corrections.
 
The principal $\rho_H,a_H$ decay modes are to lighter states with $\bar TT$
content, namely, the longitudinally-polarized $V_L = W_L,Z_L$ and the Higgs
boson~$H$.\footnote{If $\rho_H$ is coupled to third-generation quarks by an
  ETC interaction of strength $\CO(G_2)$,
  Eqs.~(\ref{eq:Gequal},\ref{eq:gapb}) imply the resulting decay rate to
  $\bar tt$ or $\bar bt$ is suppressed by the tiny factor
  $(M_{\rho_H}/\Lambda)^4$. This is much smaller than the small
  $\CO(g^4/g_{\rho_H}^2)$ $\bar tt$-rate induced by mixing with the EW
  bosons.} The two-body decays allowed by parity and isospin are
\bea\label
{eq:rhodecay}
&& \rho_H^0 \to W^+_L W^-_L,\quad \rho_H^\pm \to W_L^\pm Z_L\,;\\
\label{eq:adecay}
&& a_H^0 \to Z_L H,\quad a_H^\pm \to W_L^\pm H.
\eea
There is no allowed $\rho_H^0 \to Z_L Z_L$. The fact that $V_L,H$ also
contain third generation quarks may deplete somewhat the $\rho_H,a_H$
couplings to them. This does not alter the major decay modes in
Eqs.~(\ref{eq:rhodecay},\ref{eq:adecay}) nor affect their production rates at
the LHC. In the absence of significant depletion, the relevant $\rho_H,a_H$
couplings are induced by their $\CO(g M^2_{\rho_H}/g_{\rho_H})$ mixing with
the EW gauge bosons~\cite{Lane:2009ct,Lane:2015fza}. They are
\bea
\label{eq:rhoVV}
\CL(\rho_H \to VV) &=& -\frac{ig^2 g_{\rho_H}v^2}{2M^2_{\rho_H}}\rho^0_{H\mu\nu}
W^+_\mu W^-_\nu -\frac{ig^2 g_{\rho_H}v^2}{2M^2_{\rho_H}\cos\theta_W}
   \left(\rho^+_{H\mu\nu} W^-_\mu - \rho^-_{H\mu\nu} W^+_\mu\right)Z_\nu\,;
   \,\,\,\,\\
\label{eq:aVH}
\CL(a_H \to V H) &=& gg_{\rho_H}v
\left(a^+_{H\mu}W^-_{\mu} + a^-_{H\mu}W^+_{\mu}\right) H
+ \frac{gg_{\rho_H}v}{\cos\theta_W}a^0_{H\mu}\, Z_\mu H\,;\\
\label{eq:aVV}
\CL(a_H \to VV) &=& \frac{ig^2 g_{\rho_H}v^2}{2M^2_{\rho_H}}
a^0_{H\mu}\bigl(W^+_{\mu\nu} W^-_\nu - W^-_{\mu\nu} W^+_\nu\bigr)\nn\\
&-& \frac{ig^2 g_{\rho_H}v^2}{2M^2_{\rho_H} \cos\theta_W}
\bigl[a^+_{H\mu}\bigl(W^-_\nu Z_{\mu\nu} - W_{\mu\nu}^- Z_\nu\bigr) - {\rm
  h.c.}\bigr] ,
\eea
where $G_{\mu\nu} = \partial_\mu G_\nu - \partial_\nu G_\mu$, $g$ is the
weak-$SU(2)$ coupling, and $v=246\,\gev$. For $M_{\rho_H,a_H} \gg M_{V,H}$,
the decay rates implied by these interactions overwhelmingly involve $V_L$
and are $\CO(g^0)$: 
\bea
\label{eq:rhogamma}
\Gamma(\rho_H^0 \to W^+ W^-) &\cong& \Gamma(\rho_H^\pm \to W^\pm Z) \cong
 \frac{g_{\rho_H}^2 M_{\rho_H}}{48\pi}\,;\\
\label{eq:agamma}
\Gamma(a^0 \to ZH) &\cong& \Gamma(a^\pm \to W^\pm H) \cong
 \frac{g_{\rho_H}^2 M_{a_H}}{48\pi}\,;\\
\label{eq:gamaVV}
\Gamma(a_H^0 \to W^+ W^-) &\cong& \Gamma(a_H^\pm \to W^\pm Z) \cong
\frac{g_{\rho_H}^2 M_W^2 M_{a_H}^3}{24\pi M_{\rho_H}^4}.
\eea
The decay rates $\Gamma(\rho_H \to VV)$ and $\Gamma(a_H \to VH)$ are nearly
identical because, as explained above, $M_{\rho_H} \cong M_{a_H}$ and
$(H,{\bs V}_L)$ are an approximately degenerate $(2,2)$ quartet in the
Wigner-Weyl mode of the symmetry~\cite{Appelquist:2015vdl}. In
Eq.~(\ref{eq:gamaVV}), $a_H \to V_L V_T$, hence the
$M_W^2/M_{\rho_H}^2 = \CO(g^2/g_{\rho_H}^2)$ suppression of that rate. The
decay rate of $\rho_H^0 \to Z_L Z_T$ is similarly suppressed.

The $\rho_H$ and $a_H$ are produced at the LHC mainly by the Drell-Yan (DY)
mechanism of $\bar qq$ annihilation. The $\rho_H$ and $a_H$ have only very
weak direct coupling to light quarks, induced by ETC. Thus, their DY
production also proceeds through their mixing with the electroweak
bosons.\footnote{By a slight abuse of language, we shall refer to the HLS
  gauge bosons and the corresponding mass-eigenstates as $\rho_H$ and $a_H$.}
A secondary source of $\rho_H$ production is weak-vector boson fusion
(VBF). This VBF is dominated by $V_L V_L$ fusion and, so, it is negligibly
small for $a_H$ production in our model. This will be important in
distinguishing the nearly degenerate $\rho_H$ and $a_H$ from each other.

CMS~\cite{Khachatryan:2014gha,Khachatryan:2014hpa, Khachatryan:2016yji} and
ATLAS~\cite{Aad:2015owa,Aad:2015ufa, Aad:2015yza} reported studies of
highly-boosted $VV$ and $VH$ pairs in their Run~1 data at 8~TeV. Both
collaborations observed resonance-like excesses of 2--$3\,\sigma$ at
1.8--2.0~TeV, near the upper end of the mass range at which we would expect
to find $\rho_H \to VV$ and $a_H \to VH$. These excesses were discussed and
their significances and production rates estimated from the Run~1 data in
Ref.~\cite{Brehmer:2015dan}.

 \begin{table}[!t]
     \begin{center}{
  \begin{tabular}{|c|c|c|c|}
  \hline
 $M_{\rho_H}$ (GeV) & $\Gamma(\rho_H \to VV)$ (GeV) & $\Gamma(a_H \to VH)$  (GeV)&
 $\Gamma(a_H \to VV)$ (GeV) \\
  \hline\hline
 1800 & 178  & 184 & 0.82 \\ 
 1900 & 188  & 196 & 0.78 \\
 2000 & 198  & 208 & 0.74 \\
 \hline
 \end{tabular}}
 \caption{Principal decay rates of the isovector bosons $\rho_H$ and $a_H$ for
   $g_{\rho_H} = 3.862$ and $M_{a_H} = 1.05 M_{\rho_H}$; from
   Ref.~\cite{Lane:2015fza}. These widths may be reduced by a factor of two or
   so by the $\bar tt$ and $\bar bt$ content of $H,W_L,Z_L$. Because the
   diboson decay modes are still dominant, this does not affect their
   production rates.\label{tab:Decays}}
 \end{center}
 \end{table}
\begin{table}[!h]
     \begin{center}{
  \begin{tabular}{|c|c|c|c|c|c|}
  \hline
$\sqrt{s}$ & $M_{\rho_H}$ (GeV)& $\sigma(\rho_H^\pm)_{DY+VBF}$ (fb) & 
$\sigma(\rho_H^0)_{DY+VBF}$  (fb)& $\sigma(a_H^\pm)$ (fb) & $\sigma(a_H^0)$ (fb)\\
  \hline\hline
8 & 1800 & 1.53 $+$ 0.36 & 0.74 $+$ 0.18  & 0.71 & 0.37  \\ 
8 & 1900 & 1.05 $+$ 0.24 & 0.50 $+$ 0.12  & 0.51 & 0.27  \\
8 & 2000 & 0.73 $+$ 0.15 & 0.36 $+$ 0.075 & 0.36 & 0.17  \\
\hline\hline
13 & 1800 & 7.61 $+$ 3.67 & 3.74 $+$ 1.93  & 4.65 & 2.23  \\ 
13 & 1900 & 5.74 $+$ 2.62 & 2.81 $+$ 1.37  & 3.16 & 1.69  \\
13 & 2000 & 4.37 $+$ 1.90 & 2.16 $+$ 0.99  & 2.39 & 1.27  \\
\hline
 \end{tabular}}
 \caption{Production cross sections at the LHC of the isovector bosons
   $\rho_H$ and $a_H$ for $g_{\rho_H} = 3.862$ and $M_{a_H} = 1.05
   M_{\rho_H}$ ($\rho_H^\pm = \rho_H^+ + \rho_H^-)$. The individual DY $+$
   VBF contributions are given for $\rho_H$; the VBF rates for $a_H$ are very
   small and not given. For $g_{\rho_H} = 2.73$, $\sigma(\rho_H \to VV)$ is 
   50\% larger, $\sigma(a_H \to VH)$ is doubled, and their widths are half as
   large as in Table~2. No $K$-factor has been applied to the cross sections;
   from Ref.~\cite{Lane:2015fza}. \label{tab:Xsections}}
 \end{center}
 \end{table}

 We proposed in Ref.~\cite{Lane:2015fza} that these excesses are due to
 production of the $\rho_H$ and $a_H$ modes in
 Eqs.~(\ref{eq:rhodecay},\ref{eq:adecay}). The decay rates and cross sections
 for $M_{\rho_H} = 1.8$--2.0~TeV, $M_{a_H} = 1.05 M_{\rho_H}$ and
 $g_{\rho_H} = 1.9\,\tev/2v = 3.862$ (i.e.,
 $M_{\rho_H} \simeq \thalf g_{\rho_H} (4v)$) are given in Tables~2
 and~3.\footnote{The Drell-Yan rates in Table~3 were
   calculated using the couplings of Ref.~\cite{Lane:2002sm}, appropriate to
   a single fermion doublet, for which we assume electric charges
   $\pm \thalf$. The DY cross sections given in Ref.~\cite{Lane:2002sm} are
   easily modified for the case at hand in which there are no other light
   PGBs. They are encoded in {\sc Pythia}~6.4~\cite{Sjostrand:2006za}. Cross
   sections for $(Q_U,Q_D) = (0,-1)$ differ only slightly from these and give
   no contribution to the $H \to \gamma\gamma$ rate via a~$U$-loop.} The
 diboson resonance cross sections in Table~3 for $\sqrt{s} = 8\,\tev$ are
 $\sim 2$--10~times smaller than those estimated in Ref.~\cite{Brehmer:2015dan}
 from ATLAS and CMS Run~1 data; see Fig.~\ref{fig:TCETC_H_Fig3}. On the other
 hand, they are typical of what would be expected for Drell-Yan rates for
 $\simeq 2\,\tev$ $\rho$-like and $W'$$/Z'$ bosons decaying to dibosons. We
 shall have more to say on this below.

 The cross sections at $\sqrt{s}
 = 13\,\tev$ are 5--7 times larger than at
 $8\,\tev$.
 Preliminary reports of 2.6--$3.2\,\ifb$
 of 13-TeV data by CMS~\cite{CMS:2015nmz} and
 ATLAS~\cite{TheATLAScollaboration:2015msj, TheATLAScollaboration:2015ulg,
   TheATLAScollaboration:2015kmr} neither confirmed nor excluded the Run~1
 excesses. In August 2016, the LHC collaborations reported searches for the
 heavy diboson resonances using 13--$15\,\ifb$
 of data taken at 13~TeV. In the diboson mass range 1.8--$2.0\,\tev$,
 the ATLAS 95\%~CL upper limit on cross section times branching ratio was
 $\simge
 20\,\fb$ in the all-hadronic ($qqqq$)
 channels~\cite{ATLAS:2016yqq} as well as in the semileptonic channels
 $\ell\nu
 qq$ for $\ell = \mu,e$~\cite{ATLAS:2016cwq}. Its upper limit was
 5--$8\,\fb$
 in the sum of the semileptonic channels $\ellp\ellm
 qq$ and $\nu\nu qq$ (dominated, of course, by $\nu\nu
 qq$)~\cite{ATLAS:2016npe}. The early Run~2 data from CMS
 ($12.9\,\ifb$)
 put a 95\%~CL limit on the $\ell\nu
 qq$ channel that was also
 $20\,\fb$~\cite{CMS:2016pfl}.
 The ATLAS search for a $W/Z
 + H$ resonance in the $qq bb$ channel yielded a limit of $\sim
 12\,\fb$~\cite{ATLAS:2016cwq}. Results based on the current Run~2 data sets
 of $36\,\ifb$
 are appearing. In the all-hadronic modes $WW,WZ
 \to qqqq$ CMS reported 95\%~CL limits of $10\,\fb$ for $Z' \to WW$ and $W'
 \to WZ$ at $M_{W',Z'} \simeq 2\,\tev$~\cite{CMS:2017skt} and
 3--$4\,\fb$
 for $W'
 \to WH \to qqbb$ and $Z'\to ZH \to qqbb$, also at $M_{W',Z'} \simeq
 2\,\tev$~\cite{CMS:2017eme}. ATLAS has reported a search for $Z' \to
 ZH$ and $W' \to
 WH$ in these hadronic modes. Its 95\%~CL limits at $M_{W',Z'} =
 2\,\tev$ are $3\,\fb$ and
 $6\,\fb$,
 respectively~\cite{ATLAS:2017ywd}. Taking into account $W,Z$
 leptonic branching ratios where appropriate, all these limits are above, but
 beginning to close in on our predictions in Table~3. In any case, it is
 clear that the diboson excesses of Run~1 --- especially those in the
 all-hadronic channels depicted in Fig.~\ref{fig:TCETC_H_Fig3} --- were, at
 best, large up-fluctuations.

\begin{figure}[!t]
 \begin{center}
\includegraphics[width=6.5in, height=3in]{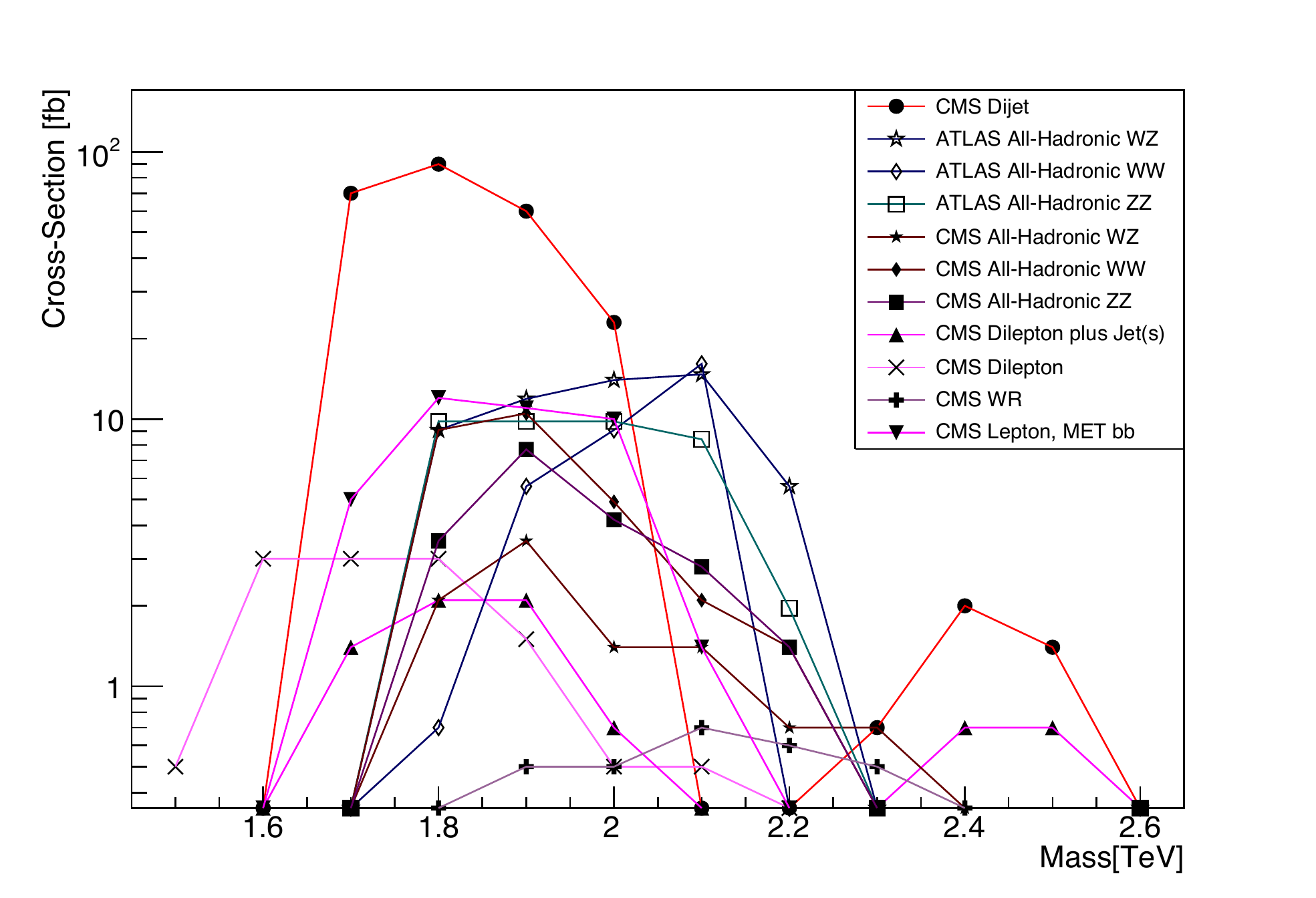}
\caption{Cross sections for ATLAS and CMS Run~1 diboson excesses estimated from
  significances greater than 1.5$\sigma$; from
  Ref.\cite{Brehmer:2015dan}. The CMS dijet with an estimated cross section
  of $\sim 100\,\fb$ was not seen in Run~2 data.}
  \label{fig:TCETC_H_Fig3}
 \end{center}
 \end{figure}

 Our proposal leads to several observations, predictions and recommendations
 for Runs~$2+3$ data analyses:

\begin{itemize}


\item[1)] Given the rather small rates in Table~3 and the low efficiency of
  separating $W$ and $Z$ in their hadronic decay modes, greater sensitivity
  to resonance signals may be had in the early Run~2 data by combining them,
  i.e., lump together the presumed $WW$, $WZ$ and $ZZ$ all-hadronic data, the
  $\ell\nu W$ and $\ell\nu Z$ data, etc.

\item[2)] The only $VV$ diboson resonances come from $\rho_H$
  production. Isospin invariance implies equal decay rates to $W^\pm Z$ and
  $W^+ W^-$, but no $ZZ$-signal.\footnote{If a $ZZ$ signal is confirmed, it
    must be due to production of another state, e.g., an analog of the
    $f_0(980)$.} It is therefore desirable that the separation of
  $p_T \sim 1\,\tev$, hadronically-decaying $W$ and $Z$-bosons be
  sharpened, and the overlap between all-hadronic $WW$, $WZ$ and $ZZ$
  selections be minimized. Until that is possible, semileptonic and
  all-leptonic $VV$ events will be needed determine the content of the
  diboson resonances. This may be feasible in Run~3 with its planned
  luminosity of $300\,\ifb$.

\item[3)] The $VH$ resonances in our model are due to $a_H$, not $\rho_H$,
  production, but they are expected to be nearly degenerate with the $VV$
  resonances. The $\rho_H$ may be distinguished by looking for forward
  jets. At $\sqrt{s} = 13\,\tev$, about 1/3 of $\rho_H \to VV$ production is
  due to VBF, which is accompanied by forward jets with a rapidity gap. The
  $a_H \to VH$ process is due entirely to DY because $a_H \to VV$ is so
  strongly suppressed and, so, it has no forward jets.

\item[4)] Table~3 shows that
  $\sigma(\rho_H^\pm \to W^\pm Z):\sigma(\rho_H^0 \to W^+W^-) \simeq
  \sigma(a_H^\pm \to W^\pm H):\sigma(a_H^0 \to ZH)\simeq 2$.
  This is a consequence of the approximate parity-doubling and the proton's
  parton luminosities at high mass. Another consequence of the parton
  luminosities is that
  $\sigma(\rho_H^+,a_H^+) \simeq 2\times \sigma(\rho_H^-,a_H^-)$.)

\item[5)] The large widths of $\rho_H$ and $a_H$ reflect their underlying
  strong dynamics, i.e., $g_{\rho_H} \simeq 3$-5, as well as to their decays
  to $V_L$. Heavy $W'$ and $Z'$ bosons are also expected to decay to
  $W^\pm_L Z_L$ and $W^+_L W^-_L$, but to be relatively narrow because their
  gauge couplings and mixings to $W,Z$ are weak. In either case, greater
  sensitivity may be obtained by detection methods favoring longitudinal
  polarization. It seems that, at least for now, the best path to diboson
  width measurements is through semileptonic $V$-decays. Again, we expect the
  same widths for $VV$ and $VH$ resonances.

\item[6)] Our model is distinguished from ones in which the composite Higgs
  is a pseudo-Goldstone boson in two ways. First, if $H$ is a PGB, there
  generally are top and $W$-partners that keep it light. They are not hadrons
  of the strong dynamics that bind $H$ and, so, are lighter than the 2-TeV
  $\rho_H$ and $a_H$. They should show up soon at the LHC. On the other hand,
  there are {\em no} top and $W$-partners needed in the strong-ETC model, and
  there aren't any. Second, the PGB models predict corrections to the $H$
  couplings with EW bosons and fermions that may be observable at the
  LHC~\cite{Brehmer:2015dan}. Such corrections in our model are suppressed by
  $(M_W/M_{\rho_H})^2$, too small to be detected at the LHC.

\end{itemize}

\section{Summary and Plans}

In this paper we developed further our strong-ETC model of electroweak
symmetry breaking and updated our discussion of the diboson resonances
$\rho_H$ and $a_H$ and their standing vis-\`a-vis the latest LHC data. We
stressed that weak-TC --- meaning a minimal role for TC in EWSB --- is a
necessary ingredient of our model if it is to explain the large mass gap
between the Higgs boson $H(125)$ and technihadrons. Our two main theoretical
purposes were to include the effect of weak TC on EWSB and to establish as
well as we could that the model provides a plausible explanation for the
lightness of $H$ relative to the technihadrons $\rho_H$ and $a_H$. For the
first, we used Takeuchi's analysis~\cite{Takeuchi:1989qa} to show that weak
TC modifies only slightly the analysis of Ref.~\cite{Lane:2014vca} in which
TC was ignored altogether. Specifically, in the $J^P = 0^\pm$ channels of
fermion-fermion scattering, the effective TC interaction has the {\em same}
form and sign as the corresponding ETC interaction, thus having the effect of
requiring only a slightly smaller ETC coupling to trigger EWSB with a light
Higgs and three Goldstone bosons $(H,{\bs V}_L)$.

Our argument that $M_{\rho_H} \gg M_H$ was an indirect one: We showed that an
{\em assumed} ETC interaction designed to generate a~$\rho_H$ pole in
technifermion scattering amplitudes leads to values of $M_{\rho_H}$ which are
either nearly zero or else much greater than $\Lambda_{TC}$ but less than
$\Lambda_{ETC}$. We have not found an argument more direct than this. We
regard both possibilities as implausible compared to
$M_{\rho_H,a_H} = \CO(\Ltc)$ --- the scale of their binding interaction ---
and significantly greater than $M_H$. Neither did we address the question of
why $\Ltc = \CO(1\,\tev)$ if TC has little to do with EWSB. Of course, we
expect that it is {\em if} the diboson excesses near $2\,\tev$ turn out to be
confirmed in LHC Runs~$2+3$. But that's not an answer. If the dibosons are
not confirmed, the question is moot. That would be unfortunate because our
model does not appear to have another readily accessible ``smoking-gun''
prediction.

On the phenomenological side, in Sec.~4 we reviewed our expectations for the
$\rho_H$ and $a_H$ widths and production cross sections~\cite{Lane:2015fza},
compared these with the latest data from Run~2, and suggested and
refined ways to search for them and distinguish our model for the diboson
resonances from others that have been proposed (reviewed in
Ref.~\cite{Brehmer:2015dan}). If they are confirmed, there will be plenty for
the experimentalists to do to reveal their nature and the interactions
responsible for them.
 
For our model, the main task remaining is to carry out a renormalization
group analysis for the Higgs and heavy fermion masses. If this analysis can
produce results in accord with experiment, particularly $M_H$ below $m_t$,
it will give strong support to our approach to understanding the Higgs as a
light composite state.

\section*{Acknowledgments}

We have benefited from conversations with T.~Appelquist, K.~Black, T.~Bose,
G.~Brooijmans, R.~Contino, E.~Eichten, B.~Holdom, A.~Martin, A.~Pomerol,
S.~Rappoccio, T.~Takeuchi, J.~Thaler and B.~Zhou. KL gratefully acknowledges
support from several sources during the course of this work: the Labex
ENIGMASS (CNRS) during 2014-16 and the Laboratoire d'Annecy-le-Vieux de
Physique Th\'eorique (LAPTh) for its continuing hospitality; the CERN Theory
Group for support and hospitality in 2014-16 and CERN for a Scientific
Associateship in 2015. This research was supported in part by the
U.S.~Department of Energy under Grant No.~DE-SC0010106.


\bibliography{TCETC_Higgs_2_final}

\providecommand{\href}[2]{#2}\begingroup\raggedright\begin{thebibliography}{10}

\bibitem{Aad:2012tfa}
{\bf ATLAS Collaboration} Collaboration, G.~Aad {\em et.~al.}, ``{Observation
  of a new particle in the search for the Standard Model Higgs boson with the
  ATLAS detector at the LHC},'' {\em Phys.Lett.} {\bf B716} (2012) 1--29,
  \href{http://xxx.lanl.gov/abs/1207.7214}{ 1207.7214}.

\bibitem{Chatrchyan:2012ufa}
{\bf CMS Collaboration} Collaboration, S.~Chatrchyan {\em et.~al.},
  ``{Observation of a new boson at a mass of 125 GeV with the CMS experiment at
  the LHC},'' {\em Phys.Lett.} {\bf B716} (2012) 30--61,
  \href{http://xxx.lanl.gov/abs/1207.7235}{ 1207.7235}.

\bibitem{Weinberg:1979bn}
S.~Weinberg, ``Implications of Dynamical Symmetry Breaking: an addendum,'' {\em
  Phys. Rev.} {\bf D19} (1979) 1277--1280.

\bibitem{Susskind:1978ms}
L.~Susskind, ``Dynamics of Spontaneous Symmetry Breaking in the Weinberg-Salam
  Theory,'' {\em Phys. Rev.} {\bf D20} (1979) 2619--2625.

\bibitem{Foadi:2012bb}
R.~Foadi, M.~T. Frandsen, and F.~Sannino, ``{125 GeV Higgs boson from a not so
  light technicolor scalar},'' {\em Phys.Rev.} {\bf D87} (2013), no.~9, 095001,
  \href{http://xxx.lanl.gov/abs/1211.1083}{ 1211.1083}.

\bibitem{Belyaev:2013ida}
A.~Belyaev, M.~S. Brown, R.~Foadi, and M.~T. Frandsen, ``{The Technicolor Higgs
  in the Light of LHC Data},'' \href{http://xxx.lanl.gov/abs/1309.2097}{
  1309.2097}.

\bibitem{Yamawaki:1986zg}
K.~Yamawaki, M.~Bando, and K.-i. Matumoto, ``Scale Invariant Technicolor Model
  and a Technidilaton,'' {\em Phys. Rev. Lett.} {\bf 56} (1986) 1335.

\bibitem{Goldberger:2008zz}
W.~D. Goldberger, B.~Grinstein, and W.~Skiba, ``{Distinguishing the Higgs boson
  from the dilaton at the Large Hadron Collider},'' {\em Phys.Rev.Lett.} {\bf
  100} (2008) 111802, \href{http://xxx.lanl.gov/abs/0708.1463}{ 0708.1463}.

\bibitem{Appelquist:2010gy}
T.~Appelquist and Y.~Bai, ``{A Light Dilaton in Walking Gauge Theories},'' {\em
  Phys.Rev.} {\bf D82} (2010) 071701, \href{http://xxx.lanl.gov/abs/1006.4375}{
  1006.4375}.

\bibitem{Bellazzini:2012vz}
B.~Bellazzini, C.~Csaki, J.~Hubisz, J.~Serra, and J.~Terning, ``{A Higgslike
  Dilaton},'' {\em Eur.Phys.J.} {\bf C73} (2013) 2333,
  \href{http://xxx.lanl.gov/abs/1209.3299}{ 1209.3299}.

\bibitem{Contino:2010rs}
R.~Contino, ``{The Higgs as a Composite Nambu-Goldstone Boson},''
  \href{http://xxx.lanl.gov/abs/1005.4269}{ 1005.4269}.

\bibitem{Giudice:2013yca}
G.~F. Giudice, ``{Naturalness after LHC8},'' {\em PoS} {\bf EPS-HEP2013} (2013)
  163, \href{http://xxx.lanl.gov/abs/1307.7879}{ 1307.7879}.

\bibitem{Bellazzini:2014yua}
B.~Bellazzini, C.~Csaki, and J.~Serra, ``{Composite Higgses},'' {\em
  Eur.Phys.J.} {\bf C74} (2014), no.~5, 2766,
  \href{http://xxx.lanl.gov/abs/1401.2457}{ 1401.2457}.

\bibitem{Barnard:2014tla}
J.~Barnard, T.~Gherghetta, T.~S. Ray, and A.~Spray, ``{The Unnatural Composite
  Higgs},'' {\em JHEP} {\bf 1501} (2015) 067,
  \href{http://xxx.lanl.gov/abs/1409.7391}{ 1409.7391}.

\bibitem{Agashe:2014kda}
{\bf Particle Data Group} Collaboration, K.~Olive {\em et.~al.}, ``{Review of
  Particle Physics},'' {\em Chin.Phys.} {\bf C38} (2014) 090001.

\bibitem{Lane:2014vca}
K.~Lane, ``{A composite Higgs model with minimal fine-tuning: The large-$N$ and
  weak-technicolor limit},'' {\em Phys.Rev.} {\bf D90} (2014), no.~9, 095025,
  \href{http://xxx.lanl.gov/abs/1407.2270}{ 1407.2270}.

\bibitem{Appelquist:1988as}
T.~Appelquist, M.~Einhorn, T.~Takeuchi, and L.~Wijewardhana, ``{Higher Mass
  Scales and Mass Hierarchies},'' {\em Phys.Lett.} {\bf B220} (1989) 223--228.

\bibitem{Chivukula:1990bc}
R.~S. Chivukula, A.~G. Cohen, and K.~D. Lane, ``{Aspects of Dynamical
  Electroweak Symmetry Breaking},'' {\em Nucl.Phys.} {\bf B343} (1990)
  554--570.

\bibitem{Bardeen:1989ds}
W.~A. Bardeen, C.~T. Hill, and M.~Lindner, ``{Minimal Dynamical Symmetry
  Breaking of the Standard Model},'' {\em Phys.Rev.} {\bf D41} (1990) 1647.

\bibitem{Panico:2015jxa}
G.~Panico and A.~Wulzer, ``{The Composite Nambu-Goldstone Higgs},''
  \href{http://xxx.lanl.gov/abs/1506.01961}{ 1506.01961}.

\bibitem{Cheng:1973nv}
T.~P. Cheng, E.~Eichten, and L.-F. Li, ``{Higgs Phenomena in Asymptotically
  Free Gauge Theories},'' {\em Phys. Rev.} {\bf D9} (1974) 2259.

\bibitem{Eichten:1979ah}
E.~Eichten and K.~D. Lane, ``{Dynamical Breaking of Weak Interaction
  Symmetries},'' {\em Phys.Lett.} {\bf B90} (1980) 125--130.

\bibitem{Appelquist:1988fm}
T.~Appelquist, M.~Soldate, T.~Takeuchi, and L.~Wijewardhana, ``{EFFECTIVE FOUR
  FERMION INTERACTIONS AND CHIRAL SYMMETRY BREAKING},''.

\bibitem{Takeuchi:1989qa}
T.~Takeuchi, ``{Analytical and Numerical Study of the Schwinger-dyson Equation
  With Four Fermion Coupling},'' {\em Phys.Rev.} {\bf D40} (1989) 2697.

\bibitem{Lane:2015fza}
K.~Lane and L.~Pritchett, ``{Heavy Vector Partners of the Light Composite
  Higgs},'' {\em Phys. Lett.} {\bf B753} (2016) 211--214,
  \href{http://xxx.lanl.gov/abs/1507.07102}{ 1507.07102}.

\bibitem{Aad:2015owa}
{\bf ATLAS} Collaboration, G.~Aad {\em et.~al.}, ``{Search for high-mass
  diboson resonances with boson-tagged jets in proton-proton collisions at
  $\sqrt{s}$ = 8 TeV with the ATLAS detector},''
  \href{http://xxx.lanl.gov/abs/1506.00962}{ 1506.00962}.

\bibitem{Aad:2015ufa}
{\bf ATLAS} Collaboration, G.~Aad {\em et.~al.}, ``{Search for production of
  $WW/WZ$ resonances decaying to a lepton, neutrino and jets in $pp$ collisions
  at $\sqrt{s}=8$ TeV with the ATLAS detector},'' {\em Eur. Phys. J.} {\bf C75}
  (2015), no.~5, 209, \href{http://xxx.lanl.gov/abs/1503.04677}{ 1503.04677}.

\bibitem{Aad:2015yza}
{\bf ATLAS} Collaboration, G.~Aad {\em et.~al.}, ``{Search for a new resonance
  decaying to a $W$ or $Z$ boson and a Higgs boson in the $\ell \ell/ \ell \nu/
  \nu \nu + b \bar{b}$ final states with the ATLAS Detector},'' {\em Eur. Phys.
  J.} {\bf C75} (2015), no.~6, 263, \href{http://xxx.lanl.gov/abs/1503.08089}{
  1503.08089}.

\bibitem{Khachatryan:2014gha}
{\bf CMS} Collaboration, V.~Khachatryan {\em et.~al.}, ``{Search for massive
  resonances decaying into pairs of boosted bosons in semi-leptonic final
  states at $\sqrt{s} =$ 8 TeV},'' {\em JHEP} {\bf 08} (2014) 174,
  \href{http://xxx.lanl.gov/abs/1405.3447}{ 1405.3447}.

\bibitem{Khachatryan:2014hpa}
{\bf CMS} Collaboration, V.~Khachatryan {\em et.~al.}, ``{Search for massive
  resonances in dijet systems containing jets tagged as W or Z boson decays in
  pp collisions at $ \sqrt{s} $ = 8 TeV},'' {\em JHEP} {\bf 08} (2014) 173,
  \href{http://xxx.lanl.gov/abs/1405.1994}{ 1405.1994}.

\bibitem{Khachatryan:2016yji}
{\bf CMS} Collaboration, V.~Khachatryan {\em et.~al.}, ``{Search for massive WH
  resonances decaying into the $\ell \nu\mathrm{ b \bar{b} }$ final state at
  $\sqrt{s}= $ 8 TeV},'' \href{http://xxx.lanl.gov/abs/1601.06431}{
  1601.06431}.

\bibitem{Nambu:1961tp}
Y.~Nambu and G.~Jona-Lasinio, ``{Dynamical Model of Elementary Particles Based
  on an Analogy with Superconductivity. 1.},'' {\em Phys.Rev.} {\bf 122} (1961)
  345--358.

\bibitem{Nambu:1961fr}
Y.~Nambu and G.~Jona-Lasinio, ``{DYNAMICAL MODEL OF ELEMENTARY PARTICLES BASED
  ON AN ANALOGY WITH SUPERCONDUCTIVITY. II},'' {\em Phys.Rev.} {\bf 124} (1961)
  246--254.

\bibitem{Nambu:1989jt}
Y.~Nambu, ``{BOOTSTRAP SYMMETRY BREAKING IN ELECTROWEAK UNIFICATION},''.

\bibitem{Miransky:1988xi}
V.~Miransky, M.~Tanabashi, and K.~Yamawaki, ``{Dynamical Electroweak Symmetry
  Breaking with Large Anomalous Dimension and t Quark Condensate},'' {\em
  Phys.Lett.} {\bf B221} (1989) 177.

\bibitem{Miransky:1989ds}
V.~Miransky, M.~Tanabashi, and K.~Yamawaki, ``{Is the t Quark Responsible for
  the Mass of W and Z Bosons?},'' {\em Mod.Phys.Lett.} {\bf A4} (1989) 1043.

\bibitem{Dobrescu:1997nm}
B.~A. Dobrescu and C.~T. Hill, ``{Electroweak symmetry breaking via top
  condensation seesaw},'' {\em Phys.Rev.Lett.} {\bf 81} (1998) 2634--2637,
  \href{http://xxx.lanl.gov/abs/hep-ph/9712319}{ hep-ph/9712319}.

\bibitem{Chivukula:1998wd}
R.~S. Chivukula, B.~A. Dobrescu, H.~Georgi, and C.~T. Hill, ``{Top quark seesaw
  theory of electroweak symmetry breaking},'' {\em Phys.Rev.} {\bf D59} (1999)
  075003, \href{http://xxx.lanl.gov/abs/hep-ph/9809470}{ hep-ph/9809470}.

\bibitem{Fukano:2012qx}
H.~S. Fukano and K.~Tuominen, ``{A hybrid 4$^{\textrm{th}}$ generation:
  Technicolor with top-seesaw},'' {\em Phys.Rev.} {\bf D85} (2012) 095025,
  \href{http://xxx.lanl.gov/abs/1202.6296}{ 1202.6296}.

\bibitem{Fukano:2013kia}
H.~S. Fukano and K.~Tuominen, ``{126 GeV Higgs boson in the top-seesaw
  model},'' {\em JHEP} {\bf 1309} (2013) 021,
  \href{http://xxx.lanl.gov/abs/1306.0205}{ 1306.0205}.

\bibitem{Bar-Shalom:2013hda}
S.~Bar-Shalom, ``{Dynamical Origin for the 125 GeV Higgs; a Hybrid setup},''
  \href{http://xxx.lanl.gov/abs/1310.2942}{ 1310.2942}.

\bibitem{Geller:2013dla}
M.~Geller, S.~Bar-Shalom, and A.~Soni, ``{Hybrid dynamical electroweak symmetry
  breaking with heavy quarks and the 125 GeV Higgs},''
  \href{http://xxx.lanl.gov/abs/1302.2915}{ 1302.2915}.

\bibitem{DiChiara:2014gsa}
S.~Di~Chiara, R.~Foadi, and K.~Tuominen, ``{125 GeV Higgs from a
  chiral-techniquark model},'' {\em Phys.Rev.} {\bf D90} (2014), no.~11,
  115016, \href{http://xxx.lanl.gov/abs/1405.7154}{ 1405.7154}.

\bibitem{DiChiara:2014uwa}
S.~Di~Chiara, R.~Foadi, K.~Tuominen, and S.~Tahtinen, ``{Dynamical Origin of
  the Electroweak Scale and the 125 GeV Scalar},''
  \href{http://xxx.lanl.gov/abs/1412.7835}{ 1412.7835}.

\bibitem{vonGersdorff:2015fta}
G.~von Gersdorff, E.~Ponton, and R.~Rosenfeld, ``{The Dynamical Composite
  Higgs},'' \href{http://xxx.lanl.gov/abs/1502.07340}{ 1502.07340}.

\bibitem{Cohen:1988sq}
A.~G. Cohen and H.~Georgi, ``{Walking Beyond the Rainbow},'' {\em Nucl.Phys.}
  {\bf B314} (1989) 7.

\bibitem{Lane:1974he}
K.~D. Lane, ``{Asymptotic Freedom and Goldstone Realization of Chiral
  Symmetry},'' {\em Phys.Rev.} {\bf D10} (1974) 2605.

\bibitem{Lane:1991qh}
K.~D. Lane and M.~Ramana, ``{Walking technicolor signatures at hadron
  colliders},'' {\em Phys.Rev.} {\bf D44} (1991) 2678--2700.

\bibitem{Bando:1987br}
M.~Bando, T.~Kugo, and K.~Yamawaki, ``{Nonlinear Realization and Hidden Local
  Symmetries},'' {\em Phys. Rept.} {\bf 164} (1988) 217--314.

\bibitem{Casalbuoni:1995qt}
R.~Casalbuoni {\em et.~al.}, ``{Degenerate BESS Model: The possibility of a low
  energy strong electroweak sector},'' {\em Phys. Rev.} {\bf D53} (1996)
  5201--5221, \href{http://xxx.lanl.gov/abs/hep-ph/9510431}{ hep-ph/9510431}.

\bibitem{Lane:2009ct}
K.~Lane and A.~Martin, ``{An Effective Lagrangian for Low-Scale Technicolor},''
  {\em Phys. Rev.} {\bf D80} (2009) 115001,
  \href{http://xxx.lanl.gov/abs/0907.3737}{ 0907.3737}.

\bibitem{Bellazzini:2012tv}
B.~Bellazzini, C.~Csaki, J.~Hubisz, J.~Serra, and J.~Terning, ``{Composite
  Higgs Sketch},'' {\em JHEP} {\bf 11} (2012) 003,
  \href{http://xxx.lanl.gov/abs/1205.4032}{ 1205.4032}.

\bibitem{Appelquist:2015vdl}
T.~Appelquist, Y.~Bai, J.~Ingoldby, and M.~Piai, ``{Spectrum-doubled Heavy
  Vector Bosons at the LHC},'' \href{http://xxx.lanl.gov/abs/1511.05473}{
  1511.05473}.

\bibitem{Kennedy:1988sn}
D.~C. Kennedy and B.~W. Lynn, ``{Electroweak Radiative Corrections with an
  Effective Lagrangian: Four Fermion Processes},'' {\em Nucl. Phys.} {\bf B322}
  (1989) 1.

\bibitem{Peskin:1990zt}
M.~E. Peskin and T.~Takeuchi, ``A new constraint on a strongly interacting
  Higgs sector,'' {\em Phys. Rev. Lett.} {\bf 65} (1990) 964--967.

\bibitem{Golden:1990ig}
M.~Golden and L.~Randall, ``Radiative corrections to electroweak parameters in
  technicolor theories,'' {\em Nucl. Phys.} {\bf B361} (1991) 3--23.

\bibitem{Holdom:1990tc}
B.~Holdom and J.~Terning, ``Large corrections to electroweak parameters in
  technicolor theories,'' {\em Phys. Lett.} {\bf B247} (1990) 88--92.

\bibitem{Altarelli:1991fk}
G.~Altarelli, R.~Barbieri, and S.~Jadach, ``Toward a model independent analysis
  of electroweak data,'' {\em Nucl. Phys.} {\bf B369} (1992) 3--32.

\bibitem{Brehmer:2015dan}
J.~Brehmer {\em et.~al.}, ``{The Diboson Excess: Experimental Situation and
  Classification of Explanations; A Les Houches Pre-Proceeding},''
  \href{http://xxx.lanl.gov/abs/1512.04357}{ 1512.04357}.

\bibitem{Lane:2002sm}
K.~Lane and S.~Mrenna, ``{The Collider phenomenology of technihadrons in the
  technicolor straw man model},'' {\em Phys.Rev.} {\bf D67} (2003) 115011,
  \href{http://xxx.lanl.gov/abs/hep-ph/0210299}{ hep-ph/0210299}.

\bibitem{Sjostrand:2006za}
T.~Sjostrand, S.~Mrenna, and P.~Skands, ``PYTHIA 6.4 physics and manual,'' {\em
  JHEP} {\bf 05} (2006) 026, \href{http://xxx.lanl.gov/abs/hep-ph/0603175}{
  hep-ph/0603175}.

\bibitem{CMS:2015nmz}
{\bf CMS} Collaboration, ``{Search for massive resonances decaying into pairs
  of boosted W and Z bosons at $\sqrt{s}$ = 13 TeV},''.

\bibitem{TheATLAScollaboration:2015msj}
``{Search for resonances with boson-tagged jets in 3.2 fb$^{−1}$ of p p
  collisions at $\sqrt{s} = 13$ TeV collected with the ATLAS detector},''.

\bibitem{TheATLAScollaboration:2015ulg}
``{Search for new resonances decaying to a W or Z boson and a Higgs boson in
  the $\ell\ell b\bar b$, $\ell\nu b\bar b$, and $\nu\nu b\bar b$ channels in
  $pp$ collisions at $\sqrt s = 13$~TeV with the ATLAS detector},''.

\bibitem{TheATLAScollaboration:2015kmr}
``{Search for $WW/WZ$ resonance production in the $\ell\nu qq$ final state at
  $\sqrt{s}=13\,$ TeV with the ATLAS detector at the LHC},''.

\bibitem{ATLAS:2016yqq}
{\bf ATLAS} Collaboration, ``{Search for resonances with boson-tagged jets in
  15.5 fb$^{-1}$ of $pp$ collisions at $\sqrt{s} = 13$ TeV collected with the
  ATLAS detector},''.

\bibitem{ATLAS:2016cwq}
{\bf ATLAS} Collaboration, ``{Search for diboson resonance production in the
  $\ell\nu qq$ final state using $pp$ collisions at $\sqrt{s}=13$ TeV with the
  ATLAS detector at the LHC},''.

\bibitem{ATLAS:2016npe}
{\bf ATLAS} Collaboration, ``{Searches for heavy ZZ and ZW resonances in the
  llqq and vvqq final states in pp collisions at sqrt(s) = 13 TeV with the
  ATLAS detector},''.

\bibitem{CMS:2016pfl}
{\bf CMS} Collaboration, ``{Search for new resonances decaying to
  $\mathrm{WW}/\mathrm{WZ} \to \ell\nu \mathrm{qq}$},''.

\bibitem{CMS:2017skt}
{\bf CMS} Collaboration, ``{Search for massive resonances decaying into WW, WZ,
  ZZ, qW and qZ in the dijet final state at $\sqrt{s} = 13~\mathrm{TeV}$},''.

\bibitem{CMS:2017eme}
{\bf CMS} Collaboration, ``{Search for heavy resonances decaying into a vector
  boson and a Higgs boson in hadronic final states with 2016 data},''.

\bibitem{ATLAS:2017ywd}
{\bf ATLAS} Collaboration, ``{Search for Heavy Resonances Decaying to a W or Z
  Boson and a Higgs Boson in the $q\bar{q}^{(\prime)}b\bar{b}$ Final State in
  $pp$ Collisions at $\sqrt{s}$ = 13 TeV with the ATLAS Detector},''.

\end{thebibliography}\endgroup
\bibliographystyle{utcaps}
\end{document}